\documentclass[12pt]{article}
\usepackage{amsmath}
\usepackage{latexsym}
\usepackage{graphicx}
\usepackage{enumerate}
\usepackage{amssymb}
\usepackage{multirow}
\usepackage[normalem]{ulem}
\usepackage{xcolor}
\DeclareGraphicsExtensions{.jpg,.pdf,.png,.gif,.eps}
\usepackage{epsfig}
\usepackage{soul}
\usepackage{epstopdf}
\usepackage{lineno}
\def\no{\noindent}

\numberwithin{equation}{section}

\begin{document}

\no {\Large {\bf Robust Hierarchical Bayes Small Area Estimation for  the Nested Error Linear Regression Model}}

\bigskip

\bigskip

\no
{\large {\bf  Adrijo Chakraborty$^1$, Gauri Sankar Datta$^{2,3}$ and Abhyuday Mandal$^2$}}

\bigskip

{\em
\no
$^1$NORC at the University of Chicago, Bethesda, MD 20814, USA

\no
$^2$Department of Statistics, University of Georgia, Athens, GA 30602, USA
\no \\
$^3$Center for Statistical Research and Methodology, US Census Bureau
\no \\
E-mails: adrijo.chakraborty@gmail.com  , gauri@stat.uga.edu  and \ \ amandal@stat.uga.edu
}

\bigskip

\bigskip

\no
{\bf Summary}

\medskip

{\bf
\bigskip
\noindent
Standard model-based small area estimates perform poorly in presence of
outliers. Sinha and Rao (2009) developed robust frequentist predictors of small area means. In this article, we present a robust Bayesian method to handle outliers in unit-level data by extending the nested error regression model. We consider a finite mixture of normal distributions for the unit-level error to model outliers and produce noninformative Bayes predictors of small area means. Our modeling approach generalizes that of Datta and Ghosh (1991) under the normality assumption. Application of our method to a data set which is suspected to contain an outlier confirms this suspicion, correctly identifies the suspected outlier, and produces robust predictors and
posterior standard deviations of the small area means. Evaluation of several procedures including the M-quantile method of Chambers and Tzavidis (2006) via simulations shows that our proposed method is as good as other procedures in terms of bias, variability and coverage probability of confidence and credible intervals when there are no outliers. In the presence of outliers, while our method and Sinha-Rao method perform similarly, they improve over the other methods. This superior performance of our procedure shows its dual (Bayes and frequentist) dominance, which should make it  attractive to all practitioners, Bayesians and frequentists, of small area estimation.
}

\medskip

\no

\medskip

\no
{\em Disclaimer}: Any opinions and conclusions expressed herein are those of the authors
and do not necessarily reflect the views of the U.S. Census Bureau or the University of
Georgia or the NORC.

\bigskip

\bigskip

\section{Introduction}

The nested error regression (NER) model with the normality assumption for both the random effects or model error terms and the unit-level error terms has played a key role in analyzing unit-level data in small area estimation. Many popular small area estimation methods have been developed under this model. In the frequentist approach, Battese et al. (1988), Prasad and Rao (1990), and Datta and Lahiri (2000), for example, derived empirical best linear unbiased predictors (EBLUPs) of small area means. These authors used various estimation methods for the variance components and derived approximately accurate estimators of mean squared error (MSEs) of the EBLUPs.  On the other hand, Datta and Ghosh (1991) followed the hierarchical Bayesian (HB) approach to derive posterior means as HB predictors and variances of the small area means. While the underlying normality assumptions for all the random quantities are appropriate for regular data, they fail to adequately accommodate outliers. Consequently, these  frequentist/Bayesian methods are highly influenced by major outliers in the data, or break down if the outliers grossly violate distributional assumptions.

\bigskip
\noindent
Sinha and Rao (2009) investigated the robustness, or lack thereof, of the EBLUPs from the usual normal NER model in the presence of ``representative outliers''. According to Chambers (1986), a representative outlier is a ``sample element with a value that has been correctly recorded and cannot  be regarded as unique. In particular, there is no reason to assume that there are no more similar outliers in the nonsampled part of the population.'' Sinha and Rao (2009) showed via simulations for the NER model that while the EBLUPs are efficient under normality, they are very sensitive to outliers that deviate from the assumed model.

\bigskip
\noindent
To address the non-robustness issue of EBLUPs, Sinha and Rao (2009) used the $\psi$-function, Huber's Proposal 2 influence function in M-estimation, to downweight the contribution of outliers in the BLUPs and the estimators of the model parameters, both regression coefficients and variance components. Using M-estimation for robust maximum likelihood, estimators of model parameters, and robust predictors of random effects, Sinha and Rao (2009) for mixed linear models  proposed a robust EBLUP (REBLUP) of mixed effects, which they used to estimate small area means for the NER model. By using a parametric bootstrap procedure they have also developed estimators of the MSEs of the REBLUPs. We refer to Sinha and Rao (2009) for details of this method. Their simulations show that when the normality assumptions hold, the proposed REBLUPs perform similar to the EBLUPs in terms of empirical bias and empirical MSE. But, in presence of outliers in the unit-level errors, while both EBLUPs and REBLUPs remain approximately unbiased, the empirical MSEs of the EBLUPs are significantly larger than those of the REBLUPs.

\bigskip
\noindent
Datta and Ghosh (1991) proposed a noninformative HB model to predict finite population small area means. In this article we follow the approach to finite population sampling which was also followed by Datta and Ghosh (1991). Our suggested model includes the  treatment of the NER model by Datta and Ghosh (1991) as a special case. Our model facilitates accommodating outliers in the population and in the sample values. We replace the normality of the unit-level error terms by a two-component mixture of normal distributions, each component centered at zero. As in Datta and Ghosh (1991), we assume normality of  the small area effects.

\bigskip
\noindent
Simulation results of Sinha and Rao (2009) indicated that there was not enough improvement in performance of the REBLUP procedures over the EBLUPs when they considered outliers in both the unit-level error and the model error terms. To keep both analytical and computational challenges for our noninformative HB analysis manageable, we use a realistic framework and we restrict ourselves to the normality assumption for the random effects. Moreover, the assumption of zero means for the unit-level error terms is similar to the assumption made by Sinha and Rao (2009). While allowing the component of the unit-level error terms with the bigger variance to also have non-zero means to accommodate outliers might appear attractive, we note later that it is not possible to conduct a noninformative Bayesian analysis with an improper prior on the new parameter.

\bigskip
\noindent
We focus only {on} unit-level model robust small area estimation in this article. There is a substantial literature on small area estimation based on area-level data using the Fay-Herriot model (see Fay and Herriot, 1979; Prasad and Rao, 1990). The paper by Sinha and Rao (2009) also discussed robust small area estimation for an area-level model. In another paper, Lahiri and Rao (1995) discussed EBLUP and estimation of MSE under a non-normality assumption for the random effects. An early robust Bayesian approach for area-level models is due to Datta and Lahiri (1995), where they used a scale mixture of normal distributions for the random effects. It is worth mentioning that the $t$-distributions are special cases of the scale mixture of normal distributions. While Datta and Lahiri (1995) assumed long-tailed distributions for the random effects, Bell and Huang (2006) used the HB method based on the $t$ distribution, either only for the sampling errors or only for the model errors.

\bigskip
\noindent
The scale mixture of normal distributions requires specification of the mixing distribution, or in the specific case for $t$ distributions, it requires the degrees of freedom. In an attempt to avoid this specification, in a recent article Chakraborty et al. (2016) proposed a simple alternative via a two-component mixture of normal distributions in terms of the variance components for the model errors.

\section{Unit-Level HB Models for Small Area Estimation } \label{sec2}
The model-based approach to finite population sampling is very useful for  modeling unit-level data in small area estimation. The NER model of Battese et al. (1988) is a popular model for unit-level data. Suppose a finite population is partitioned into $m$ small areas, with the $i$th area having $N_i$ units. The NER model relates $Y_{ij}$, the value of a response variable $Y$ for the $j$th unit in the $i$th small area, with $x_{ij} =(x_{ij1},\cdots, x_{ijp})^T$, the value of a $p$-component covariate vector associated with that unit, through a mixed linear model given by
\begin{equation}
Y_{ij}= x_{ij}^T\beta +v_i +e_{ij}, ~ j=1,\cdots, N_i, ~i=1,\cdots, m,
\label{NER-BHF}
\end{equation}
where all the random variables $v_i$'s and $e_{ij}$'s are assumed independent. Distributions of these variables are specified by assuming that random effects $v_i\stackrel{iid}\sim N(0,\sigma_v^2)$ and unit-level errors $e_{ij}\stackrel{iid}\sim N(0,\sigma_e^2)$. Here $\beta =(\beta_1,\cdots, \beta_p)^T$ is the regression coefficient vector. We want to predict the $i$th small area finite population mean ${\bar Y}_i= N_i^{-1}\sum_{j=1}^{N_i} Y_{ij}$, $i=1,\cdots,m$. We assume that the population level model (\ref{NER-BHF}) holds for any sample from the population.

\bigskip
\noindent
Battese et al. (1988) and Prasad and Rao (1990), among others, considered noninformative sampling, where a simple random sample of size $n_i$ is selected from the $i$th small area. For notational simplicity we denote the sample by $Y_{ij}, j=1,\cdots, n_i, i=1,\cdots,m$.  To develop predictors of the small area means ${\bar Y}_i, i=1,\cdots,m$, these authors first derived, for known model parameters, the conditional distribution of the {\it unsampled} values, $Y_{ij}, j=n_i +1,\cdots, N_{i}, i=1,\cdots,m$, given the sampled values $Y_{ij}, j=1,\cdots, n_{i}, i=1,\cdots,m$. Under squared error loss, the best predictor of ${\bar Y}_i$ is its mean with respect to this conditional distribution, also known as the predictive distribution. In the frequentist approach, Battese et al. (1988) and Prasad and Rao (1990) obtained the EBLUP of $\bar Y_i$ by replacing in the conditional mean the unknown model parameters $(\beta^T,\sigma_e^2,\sigma_v^2)^T$ by their estimators using $Y_{ij}, j=1,\cdots, n_i, i=1,\cdots, m$. In the Bayesian approach, on the other hand, Datta and Ghosh (1991) developed HB predictors of  ${\bar Y}_i$ by integrating out these parameters in the conditional mean of $\bar Y_i$ with respect to their posterior density, which is derived based on a prior distribution on the parameters and the distribution of the sample $Y_{ij}, j=1,\cdots, n_i, i=1,\cdots, m$,  derived under the model (\ref{NER-BHF}).

\bigskip
\noindent
While the frequentist approach for the NER model under the distributional assumptions in (\ref{NER-BHF}) continues with accurate approximation and estimation of the MSEs of the EBLUPs, the Bayesian approach typically proceeds under some noninformative priors, and computes numerically, usually by the MCMC method, the exact posterior means and {posterior  variances of the area means} $\bar Y_i$'s. Among various noninformative priors for $\beta, \sigma_e^2, \sigma_v^2$, a popular choice is
\begin{equation}
\pi_P(\beta,\sigma_e^2,\sigma_v^2) = \frac 1{\sigma_e^2},
\label{Pop-prior-BHF}
\end{equation}
(see, for example, Datta and Ghosh, 1991).

\bigskip
\noindent
The standard NER model in (\ref{NER-BHF}) is unable to explain outlier behavior of unit-level error terms. To avoid the breakdown of EBLUPs and their MSEs in the presence of outliers, Sinha and Rao (2009) modified all estimating equations for the model parameters and random effects terms by robustifying various ``standardized residuals'' that appear in the estimating equations by using Huber's $\psi$-function, which truncates large absolute values to a certain threshold. They did not replace the working NER model in (\ref{NER-BHF}) to accommodate outliers, but they accounted for their potential impacts on the EBLUPs and estimated MSEs by downweighting large standardized residuals that appear in various estimating equations through Huber's $\psi$-function. Their approach, in the terminology of Chambers et al. (2014), may be termed {\it robust projective}, where they estimated the working model in a robust fashion and used that to project sample non-outlier behavior to the unsampled part of the model.

\bigskip
\noindent
To investigate the effectiveness of their proposal, Sinha and Rao (2009) conducted simulations based on various long-tailed distributions for the random effects and/or the unit-level error terms. In one of their simulation scenarios which is reasonably simple but useful, they used a two-component mixture of normal distributions for the unit-level error terms, with both components centered at zero but with unequal variances, and the component with the larger variance appearing with a small probability. This modifies the regular setup of the NER model with the possibility of outliers arising as a small fraction of contamination caused by the error corresponding to the larger variance component. {Simulation results in Table 2 of Sinha and Rao (2009) report that outliers in the random effect have little impact on the EBLUP. Hence we  could focus on the unit-level error only}. In this article, we incorporate this mixture distribution to modify the model in (\ref{NER-BHF}) to develop new Bayesian methods that would be robust to outliers. Our proposed population level HB model is given by

\bigskip
\noindent
Normal Mixture (NM) HB Model:
\begin{itemize}
\item[(I)] Conditional on $\beta =(\beta_1,\cdots,\beta_p)^T, v_1,\cdots, v_m, z_{ij}, j=1,\cdots, N_i, i=1,\cdots, m, p_e, \sigma_1^2, \sigma_2^2$ and $\sigma_v^2$,
$$
Y_{ij} \stackrel{ind} \sim z_{ij} N(x_{ij}^T\beta+v_i , \sigma_1^2) + (1-z_{ij}) N(x_{ij}^T\beta+v_i , \sigma_2^2), ~j=1,\cdots, N_i, i=1,\cdots, m.
$$
\item[(II)] The indicator variables $z_{ij}$'s are iid with $P(z_{ij}=1|p_e)=p_e, ~j=1,\cdots, N_i, i=1,\cdots, m,$ and are independent of $\beta =(\beta_1,\cdots,\beta_p)^T, v_1,\cdots, v_m, \sigma_1^2, \sigma_2^2$ and $\sigma_v^2$.
\item[(III)] Conditional on $\beta, z=(z_{11},\cdots, z_{1N_1},\cdots, z_{m1},\cdots, z_{mN_m})^T, p_e, \sigma_1^2, \sigma_2^2$ and $\sigma_v^2$, random small area effects $v_i\stackrel{iid}\sim N(0,\sigma_v^2)$ for $i=1,\cdots, m$.
\end{itemize}
For simplicity, we assume the contamination probability $p_e$ to remain the same for all units in all small areas. Gershunskaya (2010) proposed this mixture model for empirical Bayes point estimation of small area means. We assume independent simple random samples of size $n_1, \cdots, n_m$ from the $m$ small areas. The Simple Random Sampling results in a noninformative sample and the joint distribution of responses of the sampled units can be obtained from the NM HB model above by replacing $N_i$ by $n_i$. This marginal distribution in combination with the prior distribution provided below will yield the posterior distribution of the $v_i$'s, and of all the parameters in the model. For the informative sampling developments in small area estimation we refer to Pfeffermann and Sverchkov (2007) and Verret et al. (2015).

\bigskip
\noindent
Two components of the normal mixture distribution differ only by their variances. We will assume the variance component $\sigma_2^2$ is larger than $\sigma_1^2$ and is intended to explain any outliers in a data set. However, if a data set does not include any outliers, the two component variances $\sigma_1^2, \sigma_2^2$ may only minimally differ. In such situation, the likelihood based on the sample will include limited information to distinguish between these variance parameters, and consequently, the likelihood will also have little information about the mixing proportion $p_e$. We notice this behavior in our application to a subset of the corn data in Section \ref{sec:data}. 

\bigskip
\noindent
In this article, we carry out an objective Bayesian analysis by assigning a noninformative prior to the model parameters. In particular, we propose a noninformative prior
\begin{equation}
\pi(\beta,\sigma_1^2,\sigma_2^2,\sigma_v^2,p_e) = \frac{I(0<\sigma_1^2<\sigma_2^2 <\infty)}{{(\sigma_2^2)}^2} ,
\label{new-prior}
\end{equation}
where we have assigned an improper prior on $\beta, \sigma_v^2, \sigma_1^2, \sigma_2^2$ and a proper uniform prior on the mixing proportion $p_e$. However, subjective priors could also be assigned when such subjective information is available.  Notably, it is possible to use some other proper prior on $p_e$ that may elicit the extent of contamination to the basic model to reflect prevalence of outliers. While many such subjective priors can be reasonably modeled by a beta distribution, we use a {\it uniform distribution} from this class to reflect noninformativeness or little information about this parameter. We also use the traditional uniform priors on $\beta$ and $\sigma_v^2$. In the Supplementary materials, we explore the propriety of the posterior distribution corresponding to the improper priors in (\ref{new-prior}).

\bigskip
\noindent
{The improper prior distribution on the two variances for the mixture distribution has been carefully chosen so that the prior will yield conditionally proper distributions for each parameter given the other.  The  proper conditional densities given $\sigma_2^2$ (or $\sigma_1^2$) respectively are
\begin{equation*}
\pi(\sigma_1^2|\sigma_2^2) =  \dfrac{1}{\sigma_2^2}I(0<\sigma_1^2<\sigma_2^2), ~~~
\pi(\sigma_2^2|\sigma_1^2) =  \dfrac{\sigma_1^2}{(\sigma_2^2)^2}I(\sigma_1^2<\sigma_2^2 <\infty).  
\end{equation*}
This conditional propriety is  {\it necessary} for parameters appearing in the mixture distribution in order to ensure under suitable conditions the propriety of the posterior density resulting from the HB model.  Alternatively, if we used, $\pi(\sigma_1^2,\sigma_2^2) \propto (\sigma^2_1)^{-1}(\sigma^2_2)^{-1}$, the posterior distribution would be improper  for situations when there are no observations from the outlying distribution. Prior ($\ref{new-prior}$) can accommodate these situations. The specific prior distribution that we propose above is such that the resulting marginal densities for $\sigma_1^2$ and $\sigma_2^2$ respectively, are $\pi_{\sigma_1^2}(\sigma_1^2) =(\sigma_1^2)^{-1}$ and $\pi_{\sigma_2^2}(\sigma_2^2) =(\sigma_2^2)^{-1}$. These two densities are of the same form as that of $\sigma_e^2$ in the regular model in (\ref{Pop-prior-BHF}) introduced earlier. Indeed by setting $p_e=0 $ or $1$ in our analysis, we can reproduce the HB analysis of the regular model given by (\ref{NER-BHF}) and (\ref{Pop-prior-BHF}).}

\bigskip
\noindent
We use the NM HB Model under noninformative sampling and the noninformative priors given by (\ref{new-prior}) to derive the posterior predictive distribution of $\bar{Y}_i, i=1,\cdots, m$. The NM HB model and noninformative sampling that we propose here facilitate building model for {\it representative outliers} (Chambers, 1986). According to Chambers, a representative outlier is a value of a sampled unit which is not regarded as unique in the population, and one can expect existence of similar values in the non-sampled part of the population which will influence the value of the finite population means $\bar Y_i$'s or the other parameters involved in the superpopulation model. 

\bigskip
\noindent
Following the practice of Battese et al. (1988) and Prasad and Rao (1990), we approximated the predictand $\bar Y_i$ by $\theta_i =\bar X_i^T\beta +v_i$ to draw inference on the finite population small area means. Here $\bar X_i = N_i^{-1}\sum_{j=1}^{N_i} x_{ij}$ is assumed known. This approximation works well for small sampling fractions $n_i/N_i$ and large $N_i$'s.  It has been noted by these authors, and by Sinha and Rao (2009), that even for the case of outliers in the sample the difference between the inference results  for $\bar Y_i$ and $\theta_i$ is negligible. Our own simulations for our model also confirm that {observation}. Once MCMC samples from the posterior distribution of $\beta, v_i$'s and $\sigma_v^2, \sigma_1^2, \sigma_2^2, p_e$ have been generated, using the NM HB Model the MCMC samples of $Y_{ij}, j=n_i+1,\cdots, N_i, i=1,\cdots, m$ from their posterior predictive distributions can be easily generated. Finally, using the relation $\bar Y_i =N_i^{-1}[\sum_{j=1}^{n_i} y_{ij} + \sum_{j=n_i +1}^{N_i} Y_{ij}]$, (posterior predictive) MCMC samples for $\bar Y_i$'s can be easily generated for inference on these quantities. In our own data anaysis, where the sampling fractions are negligible, we do inference for the approximated predictands $\theta_i$'s.

\bigskip
\noindent
Chambers and Tzavidis (2006) took a new frequentist approach to small area estimation that is different from the mixed model prediction used in EBLUP. Instead of using a mixed model for the response, they suggeted a method based on quantile regression. We briefly review their M-quantile small area estimation method in Section \ref{MQ-rev}. They also proposed an estimator of  MSE of their point estimators.

\bigskip
\noindent
Our Bayesian proposal has two advantages over the REBLUP of Sinha and Rao (2009). First, instead of a working model for the non-outliers, we use an explicit mixture model to specify the joint distribution of responses of all the units in the population, and not only the non-outliers part of the population. It enables us to use all the sampled observations to predict the entire non-sampled part, consisting of outliers and non-outliers, of the population. Our method is robust predictive and the noninformative HB predictors are less susceptible to bias. Second, the main thrust of the EBLUP approach in small area estimation is to develop accurate approximations and estimation of MSEs of EBLUPs (cf. Prasad and Rao, 1990). Datta and Lahiri (2000) and Datta et al. (2005) termed this approximation as second-order accurate approximation, which neglects terms lower order than $m^{-1}$ in the approximation. Second-order accurate approximation results for REBLUPs have not been obtained by Sinha and Rao (2009). Also, their bootstrap proposal to estimation of the MSE under the working model has not been shown to be second-order accurate. Our HB proposal does not rely on any asymptotic approximations. Analysis of the corn data set and simulation study show less uncertainty {(and better stability of this measure)} of our method compared to the M-quantile method.

\section{M-quantile Small Area Estimation}\label{MQ-rev}
Small area estimation is dominated by linear mixed effects models where the conditional mean of $Y_{ij}$, the response of the $j$th unit in the $i$th small area, is expressed as  $E(Y_{ij}|x_{ij}, v_i)=x_{ij}^T\beta + z_{ij}^Tv_i,$ where $x_{ij}$ and $z_{ij}$ are suitable known covariates,  $v_i$ is a random effects vector and $\beta$ is a common regression coefficient vector. This assumption is the building block for EBLUPs of small area means, based on suitable additional assumptions for this conditional distribution and the distribution of the random effects.  Also with suitable prior distribution on the model parameters, HB methodology for prediction of small area means is developed.

\bigskip
\noindent
As an alternative to linear regression which models $E(Y|x)$, the mean of the conditional distribution of $Y$ given covariates $x$,  quantile regression has been developed by modeling suitable quantiles of the conditional distribution of $Y$ given $x$. In particular in quantile linear regression, for $0<q<1$, the $q$th quantile $Q_q(Y|x)$ of this distribution is modeled as $Q_q(Y|x)=x^T\beta_q$, where $\beta_q$ is a suitable parameter modeling the linear quantile function. For a given quantile regression function, the quantile coefficient $q_i\in (0,1)$ of an observation $y_i$ satisfies $Q_{q_i}(Y|x_i)=y_i$. In particular, for a linear quantile function, for given $y_i, x_i$, the $q_i$ satisfies $x_i^T\beta_{q_i}=y_i$.

\bigskip
\noindent
While in the linear regression setup the regression coefficient $\beta$ is estimated from a set of data $\{y_i,x_i:i=1,\cdots, n\}$ by minimizing the sum of squared errors $\sum_{i=1}^n (y_i-x_i^T\beta)^2$ with respect to $\beta$, the quantile regression coefficient $\beta_q$ for a fixed $q\in (0,1)$ is obtained by minimizing the loss function $\sum_{i=1}^n |y_i-x_i^Tb| \{(1-q)I(y_i-x_i^T b \le 0) + q I(y_i-x_i^T b >0) \}$ with respect to $b$. Here $I(\cdot)$ is a usual indicator function.

\bigskip
\noindent
Following the idea of M-estimation in robust linear regression, Breckling and Chambers (1988) generalized quantile regression by minimizing an objective function $\sum_{i=1}^n d(|y_i-x_i^Tb|)\{(1-q)I(y_i-x_i^T b \le 0) + q I(y_i-x_i^T b >0) \}$ with respect to $b$ for some given loss function $d(\cdot)$. [Linear regression is a special case for $q=.5$ and $d(u)=u^2$.] Estimator of $\beta_q$ is obtained by solving the equation
$$
\sum_{i=1}^n \psi_q(r_{iq}) x_i =0,
$$
where $r_{iq} = y_i - x_i^T\beta_q$, $ \psi_q(r_{iq}) = \psi(s^{-1} r_{iq})\{(1-q)I(r_{iq} \le 0) +
q I(r_{iq}>0) \}$, the function $\psi(\cdot)$, known as the influence function in M-estimation, is determined by $d(\cdot)$ (actually, $\psi(u)$ is related to the derivative of $d(u)$, assuming it is differentiable). The quantity $s$ is a suitable scale factor determined from the data (cf.  Chambers and Tzavidis, 2006). In M-quantile regression, these authors suggested using $\psi(\cdot)$ as the Huber Proposal 2 influence function $\psi(u) = u I(|u| \le c) + c \mbox{ sign}(u) I(|u| > c)$, where $c$ is a given positive number bounded away from $0$.

\bigskip
\noindent
To apply the M-quantile method in small area estimation for a set of data $\{y_{ij}, x_{ij},j=1,\cdots, n_i, i=1,\cdots, m\}$, Fabrizi et al. (2012) followed Chambers and Tzavidis (2006) and suggested determining a set of $\hat{\beta}_q$ in a fine grid for $q\in (0,1)$ by solving
$$
\sum_{i=1}^m\sum_{j=1}^{n_i} \psi_q(r_{ijq})x_{ij} =0,
$$
where $r_{ijq} = y_{ij} -x_{ij}^T\hat{\beta}_q$. Fabrizi et al. (2012) defined M-quantile estimator of $\bar Y_i $
by
\begin{equation}
\hat{\bar Y}_{i,MQ} = \frac 1{N_i} [\sum_{j=1}^{n_i}y_{ij} + \sum_{j=n_i +1}^{N_i} x_{ij}^T\hat{\beta}_{\bar{q}_i}
+ (N_i-n_i)(\bar y_i - \bar x_i^T\hat{\beta}_{\bar{q}_i}) ],
\label{MQest}
\end{equation}
where $(\bar y_i, \bar x_i)$ is the sample mean of $\{(y_{ij}, x_{ij}), j=1,\cdots, n_i\}$.
Here $\bar q_i = \frac 1{n_i}\sum_{j=1}^{n_i} q_{ij}$ is the average estimated quantile coefficient of the $i$th small area, where $q_{ij}$ is obtained by solving $x_{ij}^T\hat{\beta}_{q} =y_{ij}$, based on the set $\{\hat{\beta}_q\}$ described above (if necessary, interpolation for $q$ is made to solve $x_{ij}^T\hat{\beta}_{q} =y_{ij}$ accurately). Here we suppress the dependence of $\hat{\beta}_q$ and $q_{ij}$ on the influence function $\psi(\cdot)$. For details on M-quantile small area estimators and associated estimators of MSE based on a pseudo-linearization method, we refer to Tzavidis and Chambers (2005) and Chambers et al. (2014).

\section{Robust Empirical Best Linear Unbiased Prediction}\label{sec:REBLUP}
Empirical best linear unbiased predictors (EBLUPs) of small area means, developed under normality assumptions for the random effects and the unit-level errors, play a very useful role in production of reliable model-based estimation methods. While the EBLUPs are efficient under the normality assumptions, they may be highly influenced by outliers in the data. Sinha and Rao (2009) investigated the robustness of the classical EBLUPs to the departure from normality assumptions and proposed a new class of predictors which are resistant to outliers. Their proposed robust modification of EBLUPs of small area means, which they termed robust EBLUP (REBLUP), downweight any influential observations in the data in estimating the model parameters and the random effects.

\bigskip
\noindent
Sinha and Rao (2009) considered a general linear mixed effects model with a block-diagonal variance-covariance matrix. Their model, which is sufficiently general to include the popular Fay-Herriot model and the nested error regression model as special cases, is given by
\begin{equation}
y_i= X_i\beta + Z_i v_i +e_i, i=1,\cdots, m,
\label{lme}
\end{equation}
for specified design matrices $X_i, Z_i$, random effects vector $v_i$ and unit-level error vector $e_i$  associated with the data $y_i$ from the $i$th small area. They assumed normality and independence of the random vectors $v_1,\cdots, v_m, e_1,\cdots, e_m$, where $v_i\sim N(0, G_i(\delta))$ and $e_i\sim N(0, R_i(\delta))$. Here $\delta$ includes the variance parameters associated with the model (\ref{lme}).

\bigskip
\noindent
To develop a robust predictor of a mixed effect $\mu_i= h_i^T\beta + k_i^Tv_i$, Sinha and Rao (2009) started with the well-known mixed model equations given by
\begin{equation}
\sum_{i=1}^m X_i^TR_i^{-1}(y_i-X_i\beta - Z_i v_i) =0, ~ Z_i^TR_i^{-1}(y_i-X_i\beta - Z_i v_i)- G_i^{-1}v_i =0,~i=1,\cdots, m,
\label{mixedmodeq}
\end{equation}
which are derived as estimating equations by differentiating the joint density of $y_1,\cdots, y_m,$ and $ v_1,\cdots, v_m$ with respect to $\beta$, and $v_1,\cdots, v_m$ to obtain``maximum likelihood'' estimators of $\beta, v_1,\cdots, v_m$ for known $\delta$. The unique solution $\tilde\beta(\delta),\tilde{v}_1(\delta),\cdots, \tilde{v}_m(\delta)$ to these equations leads to the BLUP $h_i^T\tilde\beta + k_i^T\tilde{v}_i$ of $\mu_i$. To estimate the variance parameters $\delta$, Sinha and Rao (2009) maximized the profile likelihood of $\delta$, which is the value of the likelihood of $\beta$ and $\delta$ based on the joint distribution of the data $y_1,\cdots, y_m$ at $\beta=\tilde{\beta}(\delta)$.

\bigskip
\noindent
To mitigate the impact of outliers on the estimators of the variance parameters, the regression coefficients and the random effects, Sinha and Rao (2009) extended the work of Fellner (1986) to robustify all the ``estimating equations'' by using Huber's $\psi$-function in M-estimation. Based on the robustified estimating equations, Sinha and Rao (2009) obtained the robust estimators of $\beta, \delta$ and $v_i, i=1,\cdots, m$, denoted respectively by $\hat{\beta}_M,\hat{\delta}_M$ and $\hat{v}_{iM}, i=1,\cdots, m$. These estimators lead to the REBLUP of $\mu_i$ given by $h_i^T\hat{\beta}_M + k_i^T\hat{v}_{iM}$. For details of the REBLUP and the associated parametric bootstrap estimators of the MSE of the REBLUPs of $\mu_i$, we refer the readers to the paper by Sinha and Rao (2009).

\section{Data Analysis}\label{sec:data}
We illustrate our method by analyzing the crop areas data {reported} by Battese et al. (1988) who considered EBLUP prediction of county crop areas for 12 counties in Iowa. Based on U.S. farm survey data in conjunction with LANDSAT satellite data they developed predictors of county means of hectares of corn and soybeans. Battese et al. (1988) were the first to put forward the nested error regression model for the prediction of the county crop areas. Datta and Ghosh (1991) later used the HB prediction approach on this data to illustrate Bayesian treatment of the nested error regression model. In the USDA farm survey data on 37 sampled segments from these 12 counties, Battese et al. (1988) determined in their reported data that the second observation for corn in Hardin county was an outlier so that this outlier would not unduly affect the model-based estimates of the small area means, Battese et al. (1988) initially recommended, and Datta and Ghosh (1991) subsequently followed, to remove this suspected outlier observation from their analyses. Discarding this observation results in a better fit for the nested error regression model. However, removing any data which may not be a non-representative outlier from analysis will result in loss of valuable information about a part of the non-sampled units of the population which may contain outliers.

\begin{table}[ht]\caption{Various point estimates and standard errors of county hectares of corn }\label{tab1:BHFanalysis}
	\tiny
	\begin{center}
		\begin{tabular}{|l|crr|rr|rr|rr|crr|rr|rr|rr|}
			\hline
			SA & \multicolumn{9}{|c|}{Full Data} & \multicolumn{9}{|c|}{Reduced Data}        \\
			& {$n_i$}  & \multicolumn{2}{c}{DG HB}   &\multicolumn{2}{c}{NM HB}    & \multicolumn{2}{c}{SR}  & \multicolumn{2}{c|}{MQ}& {$n_i$}  & \multicolumn{2}{c}{DG HB}   &\multicolumn{2}{c}{NM HB}    & \multicolumn{2}{c}{SR}  & \multicolumn{2}{c|}{MQ}\\
			& & Mean & SD &     Mean & SD &    Mean & SD &   Mean & SD  & & Mean & SD &     Mean & SD &    Mean & SD &   Mean & SD     \\
			\hline
			1    & 1 &    123.8 & 11.7     &    123.4 & 9.8      &   123.7 & 9.9     &   130.0 &  5.7   & 1 & 122.0 & 11.6   & 121.7 & 9.7     & 122.2 & 9.9  & 128.0 & 3.7   \\
			2    & 1 &    124.9 & 11.4     &    126.6 & 10.3     &   125.3 & 9.7     &   134.2 &  8.4   & 1 & 126.4 & 10.9   & 127.2 & 9.7     & 126.5 & 9.5  & 133.4 & 6.0   \\  
			3    & 1 &    110.0 & 12.3     &    108.0 & 11.3     &   110.3 & 9.4     &    86.0 & 18.3   & 1 & 107.6 & 12.4   & 105.6 & 10.1    & 106.7 & 9.5  &  94.6 & 14.4  \\  
			4    & 2 &    114.2 & 10.7     &    112.3 & 10.2     &   114.1 & 8.8     &   114.4 &  3.4   & 2 & 108.9 & 10.5   & 108.2 & 8.7     & 111.0 & 8.3  & 113.3 & 3.7   \\  
			5    & 3 &    140.3 & 10.8     &    142.1 & 8.1      &   140.8 & 7.8     &   144.2 & 11.3   & 3 & 143.6 & 9.7    & 144.1 & 7.0     & 143.3 & 7.1  & 144.2 & 9.3   \\  
			6    & 3 &    110.0 & 9.6      &    111.4 & 7.6      &   110.8 & 7.6     &   108.6 &  3.9   & 3 & 112.3 & 9.7    & 112.5 & 6.5     & 112.3 & 7.1  & 114.5 & 5.4   \\  
			7    & 3 &    116.0 & 9.7      &    114.3 & 7.6      &   115.2 & 7.3     &   116.3 &  4.2   & 3 & 113.4 & 9.1    & 112.5 & 6.8     & 112.9 & 7.1  & 115.4 & 3.8   \\  
			8    & 3 &    123.2 & 9.5      &    122.7 & 7.9      &   122.7 & 7.5     &   122.5 &  3.9   & 3 & 121.9 & 8.8    & 121.9 & 6.6     & 121.9 & 7.1  & 122.7 & 4.0   \\  
			9    & 4 &    112.6 & 9.9      &    113.9 & 6.9      &   113.5 & 6.5     &   115.3 &  5.8   & 4 & 115.5 & 9.2    & 115.7 & 5.7     & 115.3 & 6.4  & 115.7 & 4.6   \\  
			10   & 5 &    124.4 & 8.9      &    123.5 & 6.1      &   124.1 & 6.3     &   121.6 &  4.7   & 5 & 124.8 & 8.4    & 124.4 & 5.4     & 124.5 & 5.3  & 123.1 & 4.0   \\  
			11   & 5 &    111.3 & 8.9      &    108.2 & 6.8      &   109.5 & 6.2     &   106.9 & 10.6   & 5 & 107.7 & 8.5    & 106.3 & 5.7     & 106.8 & 5.4  & 105.5 & 7.0   \\  
			12   & 6 &    130.7 & 8.3      &    135.3 & 7.5      &   136.9 & 6.0     &   135.8 &  4.3   & 5 & 142.6 & 9.0    & 143.5 & 5.9     & 143.1 & 5.8  & 140.6 & 4.9   \\  
			\hline
		\end{tabular}
	\end{center}
\end{table}

\bigskip
\noindent
We reanalyze the full data set for corn using our proposed HB method as well as the other methods we reviewed above.
In Table~\ref{tab1:BHFanalysis} we report various point estimates and standard error estimates. We compare our proposed robust HB prediction method with the standard HB method of Datta and Ghosh (1991), and two robust frequentist methods, the REBLUP method of Sinha and Rao (2009) and the MQ method of Chambers and Tzavidis (2006).
We list in the table various estimates of county hectares of corn, along with their estimated standard errors or posterior standard deviations. Our analysis of the full data set including the potential outlier from the last small area shows that for the first 11 small areas there is a close agreement among the three sets of point estimates by Datta and Ghosh (1991), Sinha and Rao (2009) and the proposed normal mixture HB method. The Datta and Ghosh method, which was not developed to handle outliers, yields a point estimate for the 12th small area that is much different from the point estimates from Sinha-Rao or the proposed NM HB method. The latter two robust estimates are very similar in terms of point estimates for all the small areas. But when we compare these two sets of robust estimates with those from another robust method, namely, the MQ estimates, we find that the MQ estimates for the first three small areas are widely different from those for the other two methods. These numbers possibly indicate a potential bias of the MQ estimates.

\bigskip
\noindent
To compare performance of all these methods in the absence of any potential outliers, we reanalyzed the corn data by removing the suspected outlier (our robust HB analysis confirmed the outlier status of this observation, cf. Figure \ref{BHF-Reduced} below). When we compare the MQ estimates with the four other sets of estimates, the DG HB, the SR, the NM, which are reported in Table  \ref{tab1:BHFanalysis}, and the EB estimates from Table 3 of Fabrizi  et al. (2012), we notice a great divide between the MQ estimates and the other estimates. Out of the twelve small areas, the estimates for areas 1, 2, 3, 5, and 6 from the MQ method differ substantially from the estimates from the other four methods. On the other hand, the close agreement among the last four sets of estimates also shows in general the usefulness of the robust predictors, the proposed HB predictors and the Sinha-Rao robust EBLUP predictors.

\bigskip
\noindent
To examine the influence of the outlier on the estimates we compare changes in the estimates from both the full  and  reduced data. We find that the largest change occurs, not surprisingly, for the DG HB method for the small area suspected of the outlier. Such a large change occurred since the DG method cannot suitably downweight an outlier, consequently, it treated the outlier value of 88.59 in the same manner as it treated any other non-outlier observation. As a result, the predictor substantially underestimated the true mean $\bar Y_i$ for  Hardin county. The next largest difference occurred for the MQ method for small area 3 which is not known to include any outlier. Such a large change is contrary to behavior of a robust method.

\begin{figure}[h]
	\begin{center}
		\begin{tabular}{cc}
			\includegraphics[scale=.35]{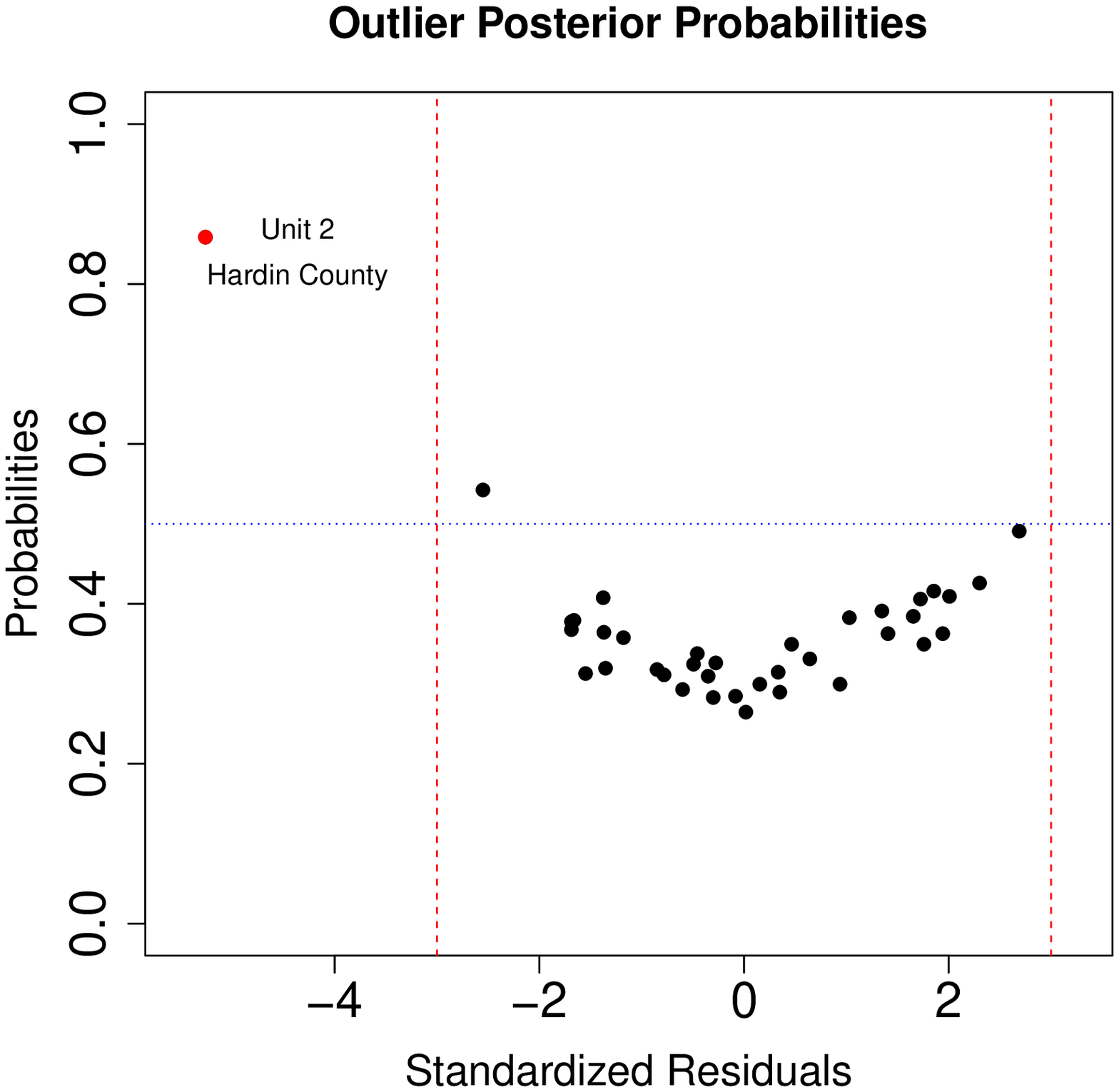} & \includegraphics[scale=.35]{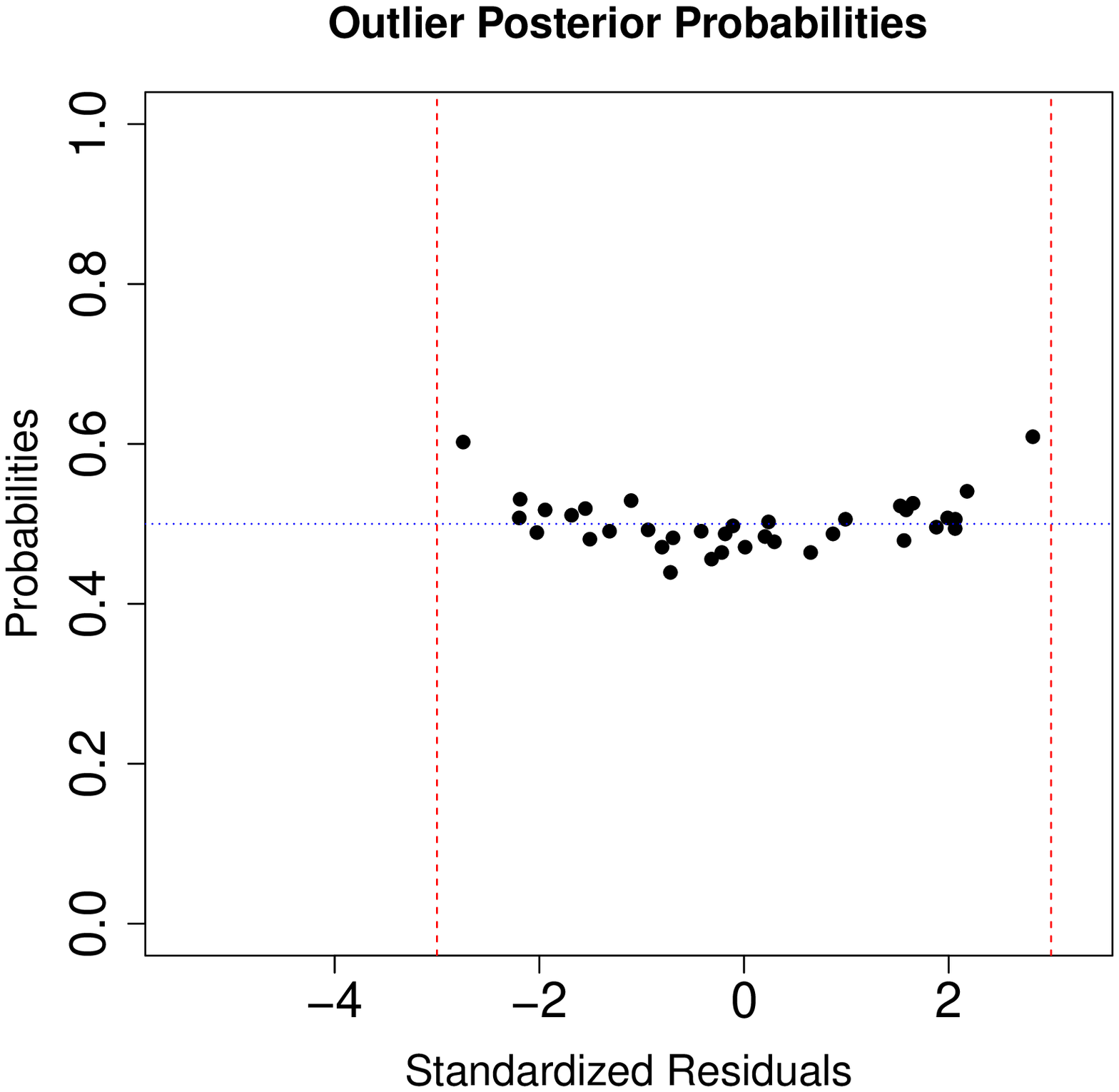}
		\end{tabular}
		\caption{Posterior probabilities of observations being outliers in {\it full} and {\it reduced} data }\label{BHF-Reduced}
	\end{center}
\end{figure}

\bigskip
\noindent
The changes in point estimates for the robust HB and the REBLUP methods are moderate for the areas not known to include any outliers, and the changes seem proportionate for the small area suspected of an outlier. The corresponding changes in the estimates from the MQ method for some of the  areas not including any outlier seem disproportionately large, and the change in the estimate for the area suspected of an outlier is not as large. This behavior to some extent indicates a lack of robustness of the MQ method to outliers.

\bigskip
\noindent
An inspection of the posterior standard deviations of the two Bayesian methods reveals some interesting points.
First, the posterior SDs of the small area means for the proposed mixture model appear to be substantially smaller than the posterior SDs associated with the Datta-Ghosh HB estimators. Smaller posterior SDs suggest the  posterior distribution of the small area means under the mixture model are more concentrated than those under the Datta-Ghosh model. This has been confirmed by simulation study, reported in the next section.

\bigskip
\noindent
Next, when we compare the posterior SDs of small area means for our proposed method based on the full data and the reduced data, all posterior SDs increase for the full data (which likely contain an outlier). In the presence of outliers, the unit-level variance is expected to be large. Even though the posterior SDs of the small area means do not depend entirely only on the unit-level error variance, they are expected to increase with this variance.  This monotonic increase appears reasonable due to the suspected outlier. While this intuitive property holds for our proposed method, it does not hold for the standard Datta-Ghosh method.

\bigskip
\noindent
For further demonstration of the effectiveness of our proposed robust HB method, we computed model parameter estimates for both the reduced and the full data sets. These estimates are displayed in Table \ref{tabMPE:BHFanalysis}. The HB estimate of the larger variance component (976, based on mean) of the mixture is much larger than the estimate of the smaller component (182) for the full data, indicating a necessity of the mixture model.  On the other hand, for the reduced data the estimates of variances for the two mixing components, 231 and 121, respectively are very similar and can be argued identical within errors in estimation, indicating limited need of the mixture distribution. A comparison of the estimates of $p_e$ for the two cases also reveals the appropriateness of the mixture model for the full data. It also shows the redundancy of including $p_e$ in the modeling of the reduced data as explained below.

\begin{table}[h]\caption{Parameter estimates for various models with and without the suspected outlier   }\label{tabMPE:BHFanalysis}


\begin{center}
\scriptsize
  \begin{tabular}{c|rr|rr|rr|rr|rr}
    \hline
    Estimates            & \multicolumn{2}{c}{Datta-Ghosh HB} & \multicolumn{2}{c}{Datta-Ghosh HB} & \multicolumn{2}{c}{Proposed Mixture HB} & \multicolumn{2}{c}{Proposed Mixture HB} & \multicolumn{2}{c}{Sinha-Rao}\\
    Estimates            & \multicolumn{2}{c}{MEAN          } & \multicolumn{2}{c}{MEDIAN        } & \multicolumn{2}{c}{MEAN               } & \multicolumn{2}{c}{MEDIAN             } & \multicolumn{2}{c}{Sinha-Rao}\\
                         & Full         & Reduced             & Full         & Reduced             & Full         & Reduced                  & Full         & Reduced                  & Full         & Reduced              \\
                         & Data         & Data                & Data         & Data                & Data         & Data                     & Data         & Data                     & Data         & Data                 \\
    \hline
    $\hat{\beta}_0$      & $ 17.29$     & $ 50.35$            & $ 16.17$     & $ 50.92$            & $ 30.89$     & $ 49.98$                 & $ 31.46$     & $ 50.78$                 & $ 29.14$     & $ 48.20$             \\
    $\hat{\beta}_1$      & $  0.37$     & $  0.33$            & $  0.37$     & $  0.33$            & $  0.35$     & $  0.33$                 & $  0.35$     & $  0.33$                 & $  0.36$     & $  0.34$             \\
    $\hat{\beta}_2$      & $ -0.03$     & $ -0.13$            & $ -0.03$     & $ -0.13$            & $ -0.07$     & $ -0.13$                 & $ -0.07$     & $ -0.13$                 & $ -0.07$     & $ -0.13$             \\
    $\hat{p}_e$          & $     -$     & $     -$            & $     -$     & $     -$            & $  0.62$     & $  0.50$                 & $  0.68$     & $  0.49$                 & $     -$     & $     -$             \\
    $\hat{\sigma}_v^2$   & $175.68$     & $231.87$            & $127.68$     & $186.07$            & $205.01$     & $238.42$                 & $160.22$     & $203.55$                 & $102.74$     & $155.15$             \\
    $\hat{\sigma}_1^2$   & $     -$     & $     -$            & $     -$     & $     -$            & $182.01$     & $121.40$                 & $170.64$     & $119.49$                 & $     -$     & $     -$             \\
    $\hat{\sigma}_2^2$   & $370.00$     & $216.00$            & $341.00$     & $192.00$            & $976.00$     & $231.00$                 & $483.00$     & $188.00$                 & $225.60$     & $161.50$             \\
  \hline
  \end{tabular}

\end{center}
\end{table}

\bigskip 
\noindent
The posterior density in a reasonable noninformative Bayesian analysis is usually dominated by the likelihood of the parameters generated by the data. In case the data do not provide much information about some parameters to the likelihood,  posterior densities of such parameters will be dominated by their prior information. Consequently,  the posterior distribution for some of them may be very similar to the prior distribution. An overparameterized likelihood usually carries little information for some parameters responsible for overparameterization.  In particular, if our mixture model is overparameterized in the sense that variances of mixture components are similar, then the integrated likelihood may be flat on the mixing proportion. We observe this scenario in our data analysis when we removed the suspected outlier observation from analysis based on our model. Since our mixture model is meant to accommodate outliers based on unequal variances for the mixing components, in the absence of any outliers the mixture of two normal distributions may not be required. In particular, we noticed earlier that with the suspected outlier removed the estimates of the two variance components $\sigma_1^2$ and $\sigma_2^2$ are very similar. Also, the posterior histogram of the mixing proportion $p_e$, not presented here, resembles a uniform distrubution, the prior distribution assigned in our Bayesian analysis. In fact, the posterior mean of this parameter for the reduced data is the same as the prior mean $0.5$. This essentially says that the likelihood is devoid of any information about $p_e$ to update the prior distribution.

\bigskip
\noindent
One advantage of our mixture model is that it explicitly models any representative outlier through the latent indicator variable $z_{ij}$. By computing the posterior probability of $z_{ij}=0$ we can compute the posterior probability that an observed $y_{ij}$ is an outlier. While the REBLUP method does not give a similar measure for an observation, one can determine the outlier status by computing the standardized residual associated with an observation. To show the effectiveness of our method, in Figure \ref{BHF-Reduced}, we plotted the posterior probabilities of an individual observation being an outlier against the observation's standardized residual. In the left panel, we showed the plot of these posterior probabilities for the full data, and  in the right panel we included the same by removing the suspected outlier. These two figures are in sharp contrast; the left panel clearly showed that there is a high probability (0.86) that the second observation in Hardin county is an outlier. The associated large negative standardized residual of this observation also confirmed that, and from this plot an approximate monotonicity of these posterior probabilities with respect to the absolute values of the standardized residuals may also be discerned. However, the right panel shows that for the reduced data excluding the suspected outlier, the standardized residuals for the remaining observations are between $-3$ and $3$, with the associated posterior probabilities of being outlier observations are all between 0.44 and 0.64. None of these probabilities is particularly larger than prior probability 0.5 to indicate outlier status of that corresponding observation. This little change of the outlier prior probabilities in the posterior distribution for the reduced data essentially confirms that a discrete scale mixture of normal distributions is not supported by the data, or in other words, the scale mixture model is not required to explain the data, which is the same as that there are possibly no outliers in the data set.

\section{A Simulation Study}\label{sec:simul}
In our extensive simulation study, we followed the simulation setup used by Sinha and Rao (2009). Corresponding to the model in (\ref{NER-BHF}), we use a single auxiliary variable $x$, which we generated independently from a normal distribution with mean $1$ and variance $1$. In our simulations we use $m=40$. We generated 40 sets of 200 ($=N_i$) values of $x$ to create the finite population of covariates for the 40 small areas.  Based on these simulated values we computed $\bar X_i = \frac 1{N_i}\sum_{j=1}^{N_i} x_{ij}$. Throughout our simulations we keep the generated $x$ values fixed. We used these generated $x_{ij}$ values and generated $v_i, i=1,\cdots, m$ independently from $N(0,\sigma_v^2)$ with $\sigma_v^2=1$. We generated $e_{ij}, j=1,\cdots, N_i, i=1,\cdots,m$ as iid from one of three possible distributions: (i) the case of no outliers where $e_{ij}$ are generated from $N(0,1)$ distribution; (ii) a mixture of normal distributions, with 10\% outliers from a $N(0,5^2)$ distribution and the remaining 90\% from the $N(0,1)$ distribution; and (iii) $e_{ij}$'s are iid from a $t$-distribution with 4 degrees of freedom. We also took $\beta_0 =1$ and $\beta_1 =1$ as in Sinha and Rao (2009), and generated $m$ small area finite populations based on the generated $x_{ij}$'s, $v_i$'s and $e_{ij}$'s by computing $Y_{ij} =\beta_0 +\beta_1 x_{ij} +v_i +e_{ij}$ based on the NER model in (\ref{NER-BHF}). Our goal is prediction of finite population small area means $\bar Y_i =\frac 1{N_i}\sum_{j=1}^{N_i} Y_{ij}, i=1,\cdots, m$. After examining no significant difference between $\bar Y_i$ and $\beta_0+\beta_1\bar X_i +v_i=\theta_i$ (say) in the simulated populations, as in Sinha and Rao (2009), we also consider prediction of $\theta_i$.

\bigskip
\noindent
From each simulated small area finite population we selected a simple random sample of size $n_i=4$ for each small area. Based on the selected samples we derived the HB predictors of Datta and Ghosh (1991) (referred to as DG), the REBLUPs of Sinha and Rao (2009) (referred to as SR), the MQ predictors of Chambers et al. (2014) (referred to  as CCST-MQ, based on their equation (38)) and our proposed robust HB predictors (referred to as NM). In addition to the point predictors we also obtained the posterior variances of both the HB predictors and the estimates of the MSE of the REBLUPs based on the bootstrap method proposed by Sinha and Rao (2009), and the estimates of MSE of the MQ predictors, obtained by using pseudo-linearization in equation (39) of Chambers et al. (2014).

\begin{figure}[h]
	\begin{center}
		\begin{tabular}{ccc}
			\raisebox{3cm}{10\%\hspace{.75cm}}       & \includegraphics[scale=.375]{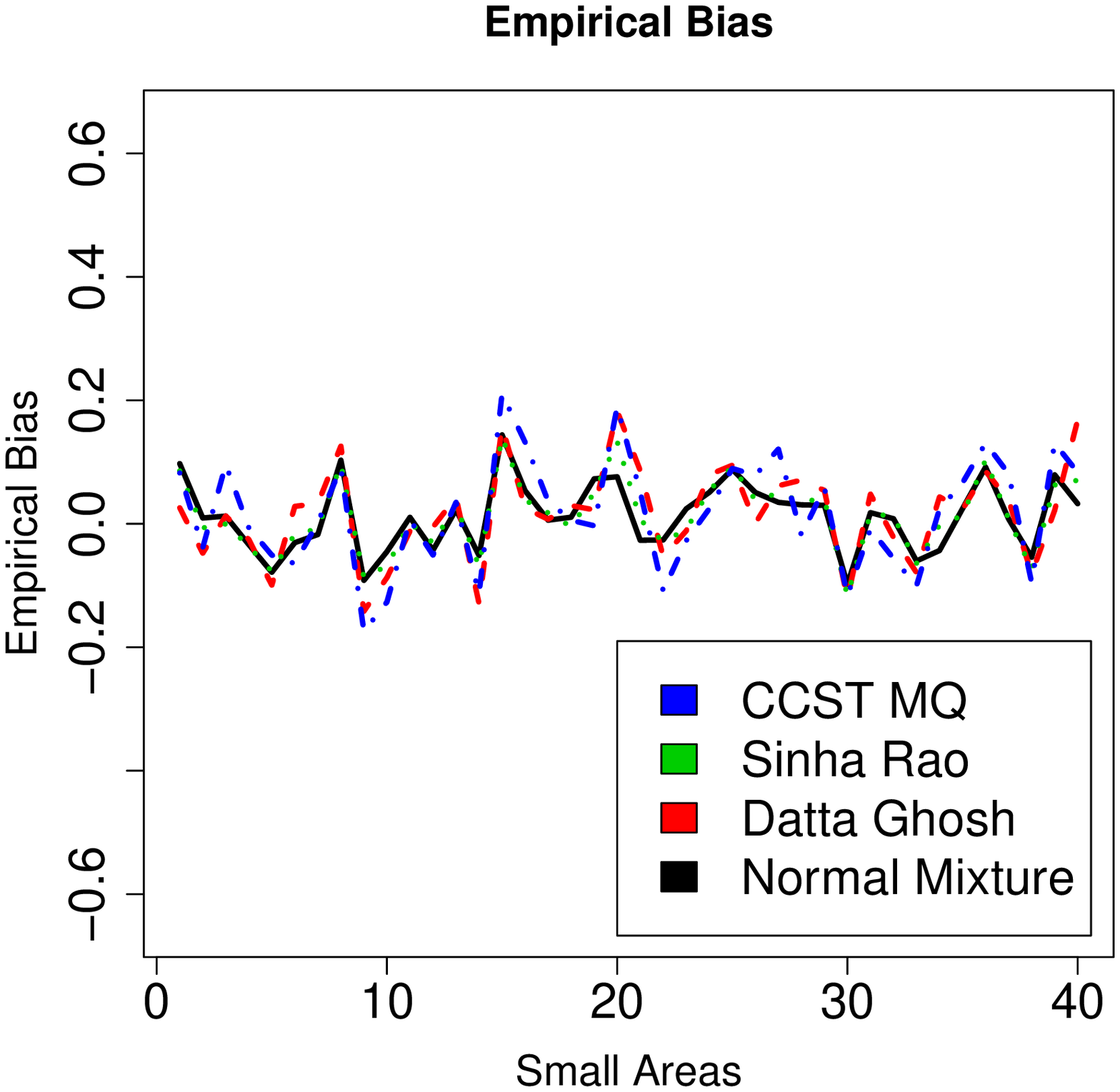} & \includegraphics[scale=.375]{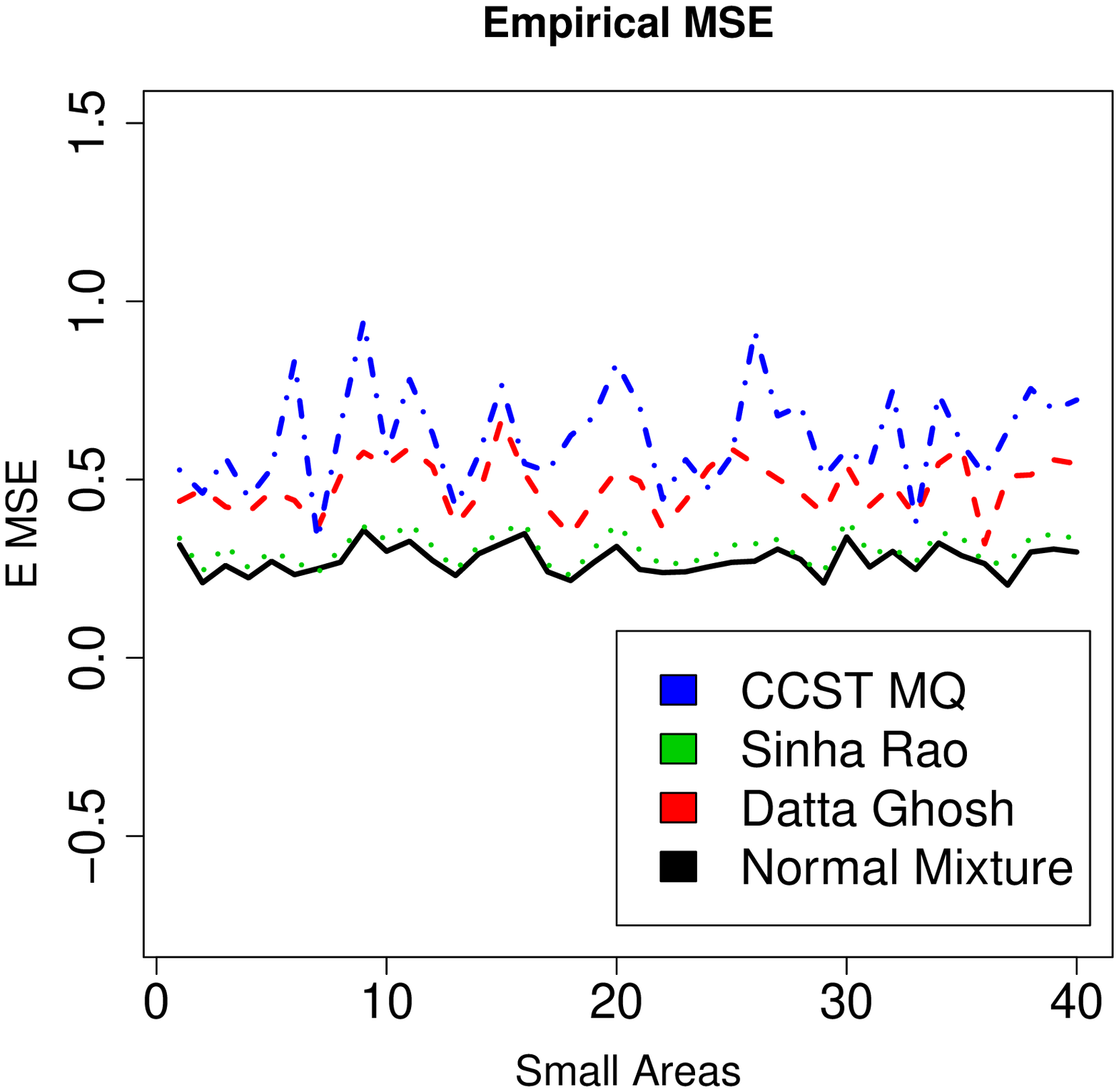} \\
			\raisebox{3cm}{$t_{(4)}$\hspace{.75cm}}  & \includegraphics[scale=.375]{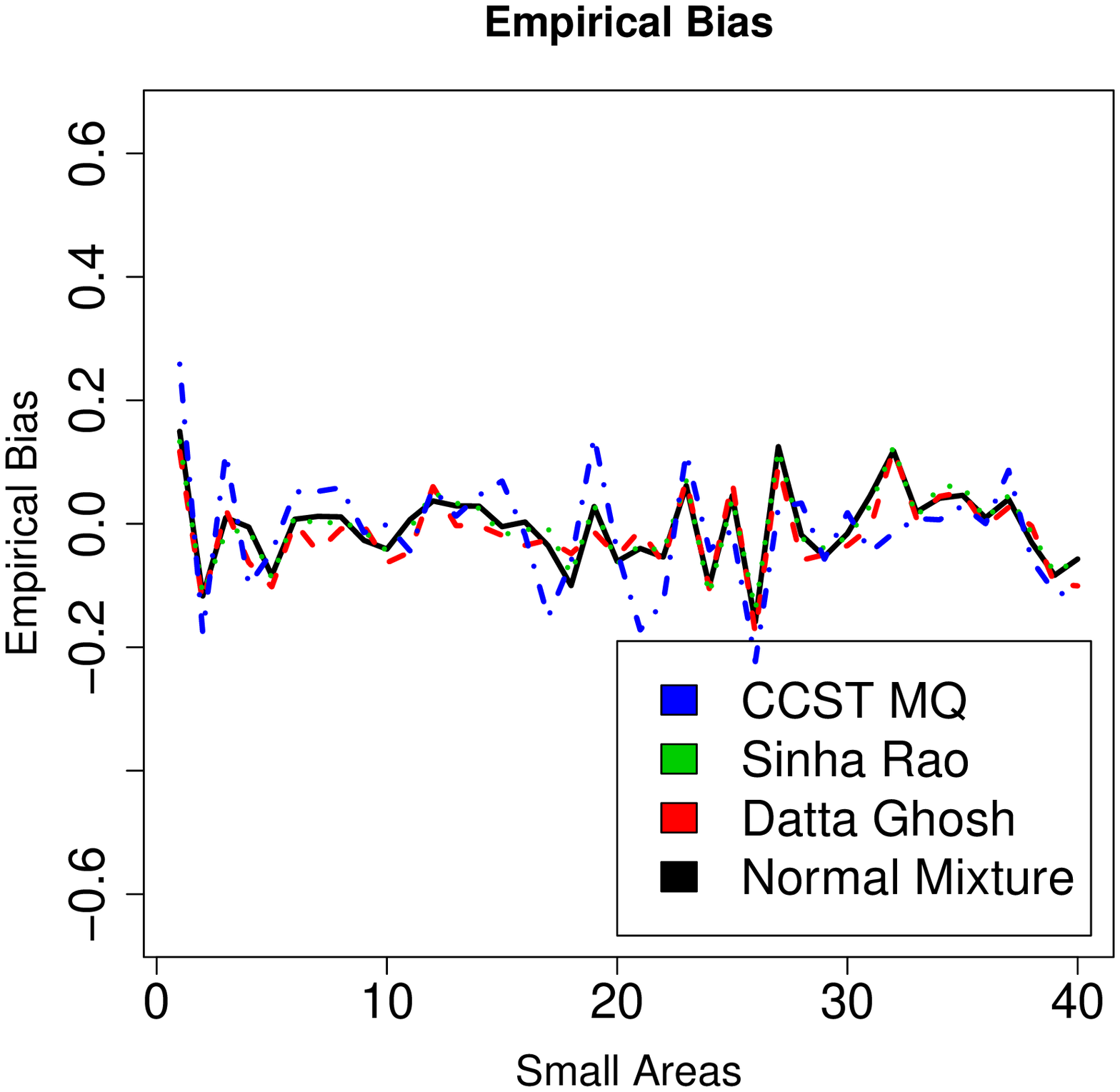}  & \includegraphics[scale=.375]{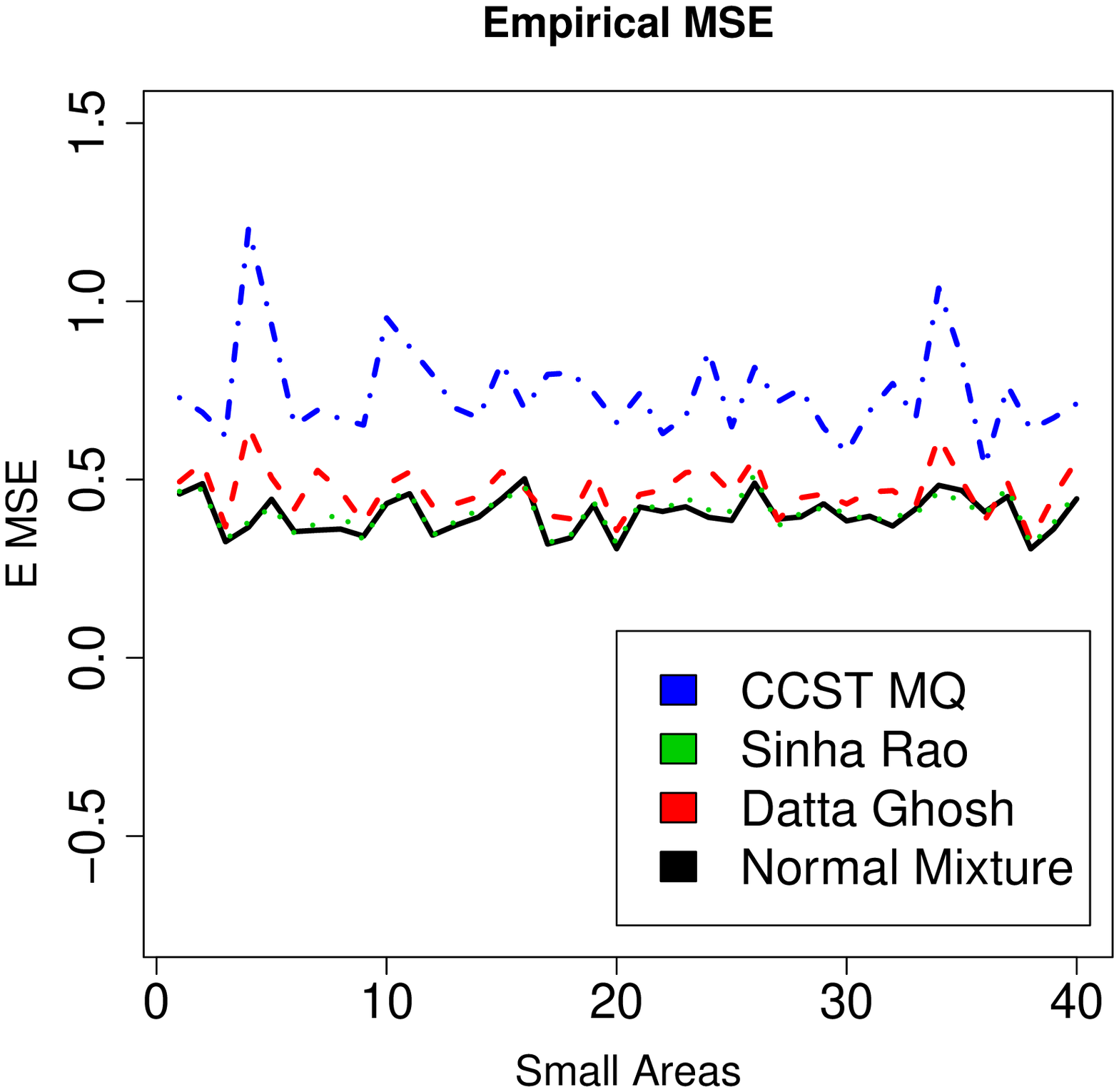}  \\
			\raisebox{3cm}{No outlier}               & \includegraphics[scale=.375]{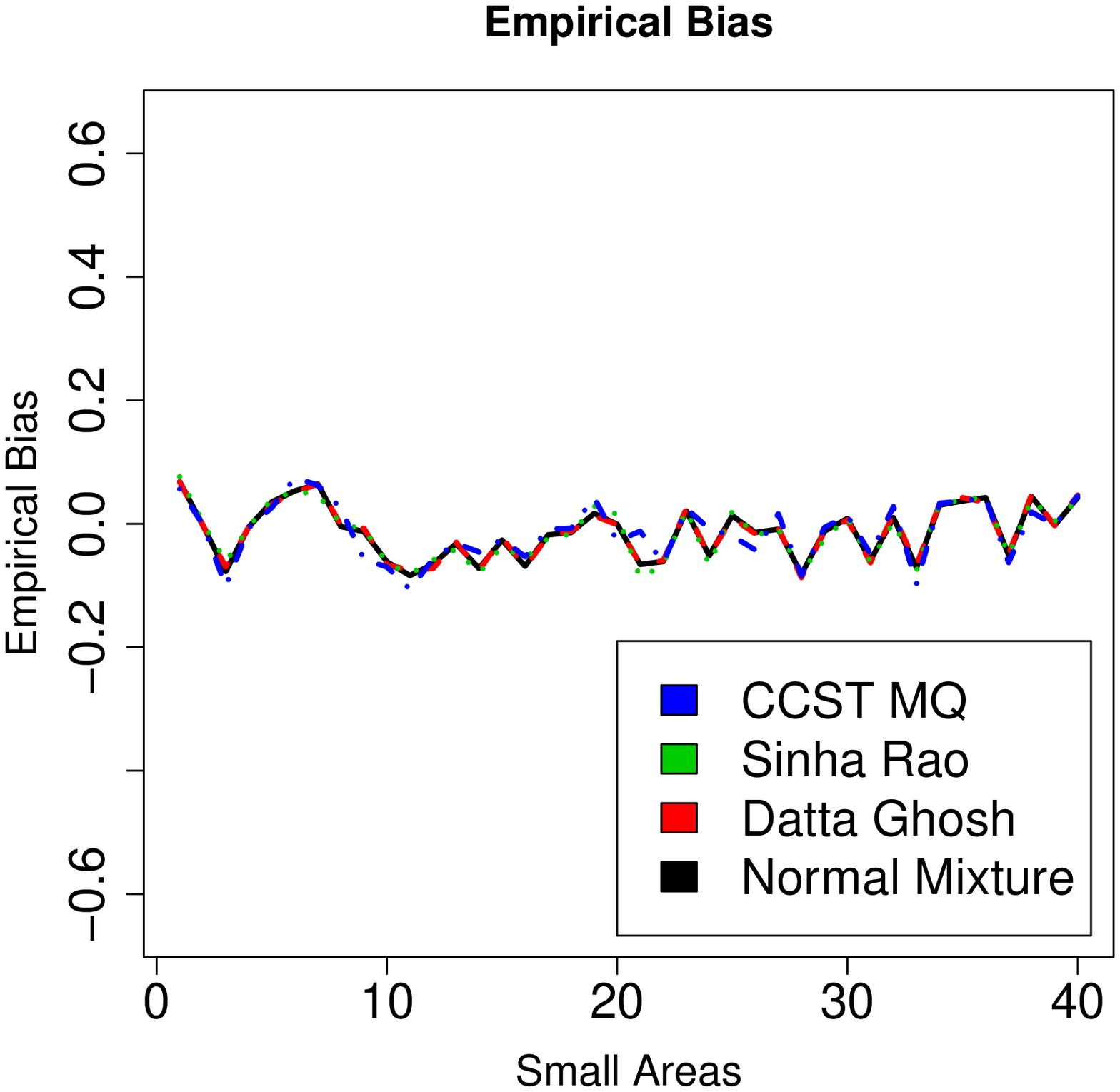} & \includegraphics[scale=.375]{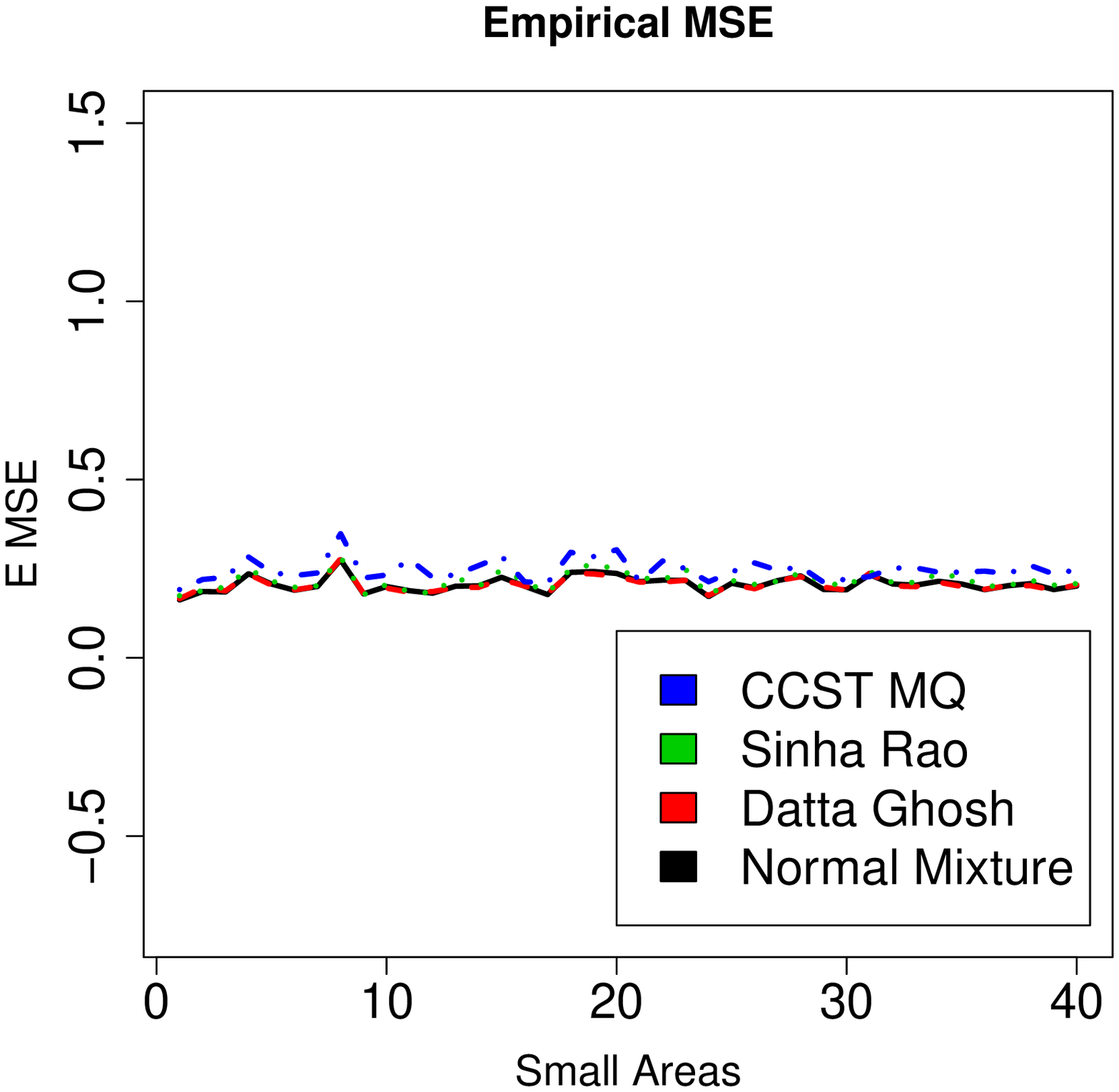}
		\end{tabular}
		\caption{Plot of empirical biases and empirical MSEs of $\hat{\theta}$s}\label{tout:BiasVar3}
	\end{center}
\end{figure}

\begin{figure}[h]
	\begin{center}
		\begin{tabular}{ccc}
			\raisebox{3cm}{10\%\hspace{.75cm}}       & \includegraphics[scale=.375] {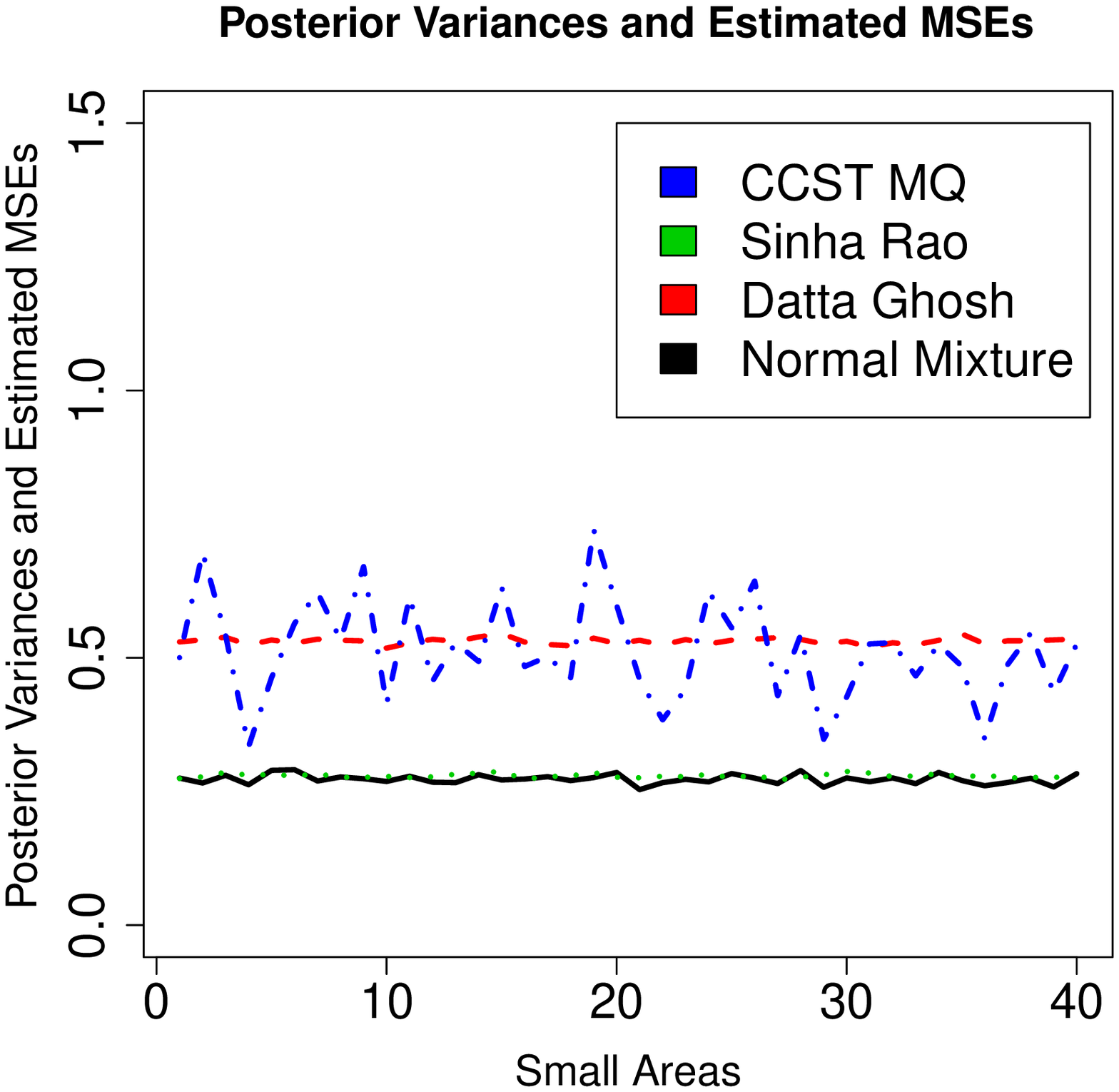}& \includegraphics[scale=.375]{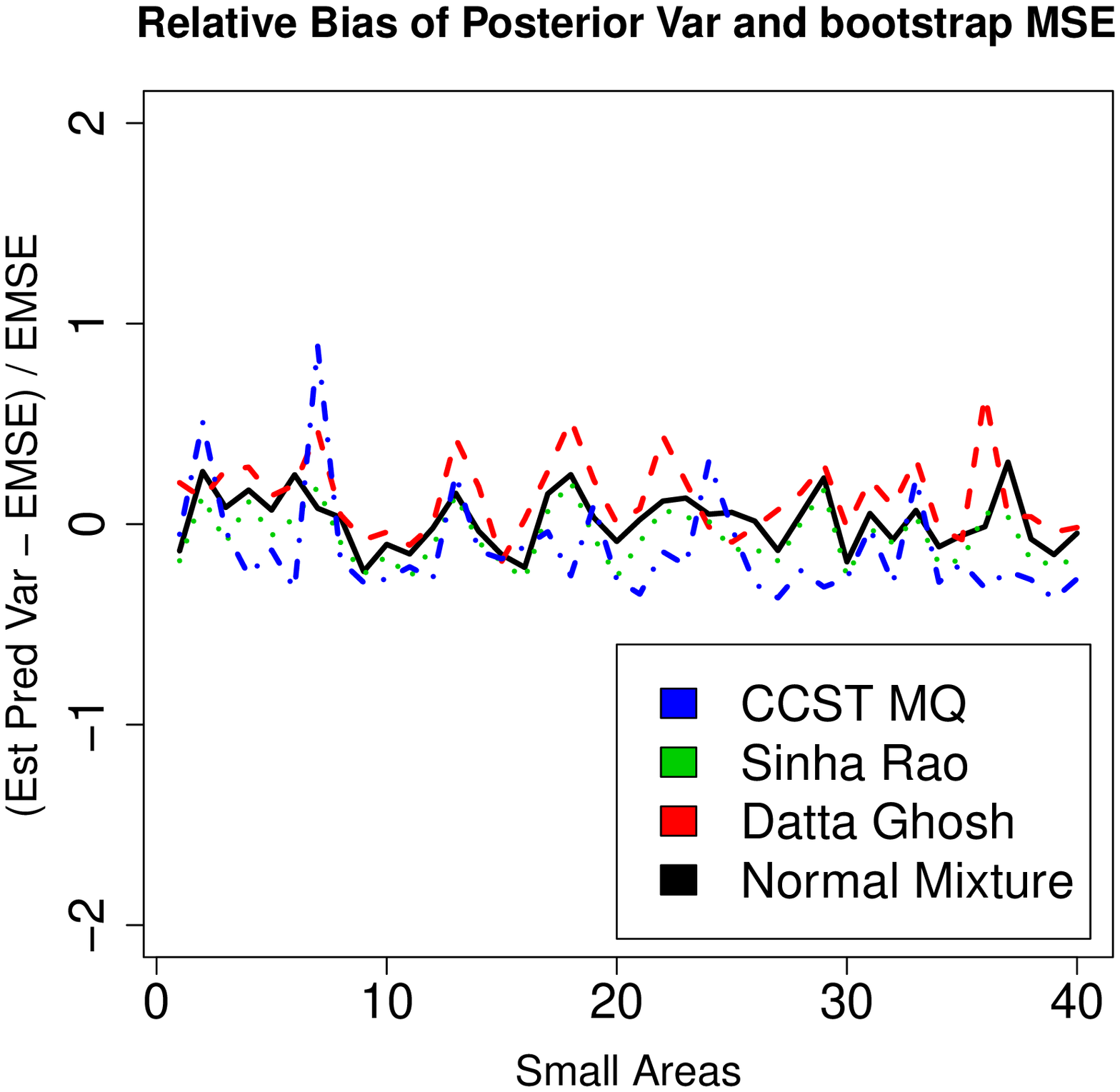}  \\
			\raisebox{3cm}{$t_{(4)}$\hspace{.75cm}}  & \includegraphics[scale=.375] {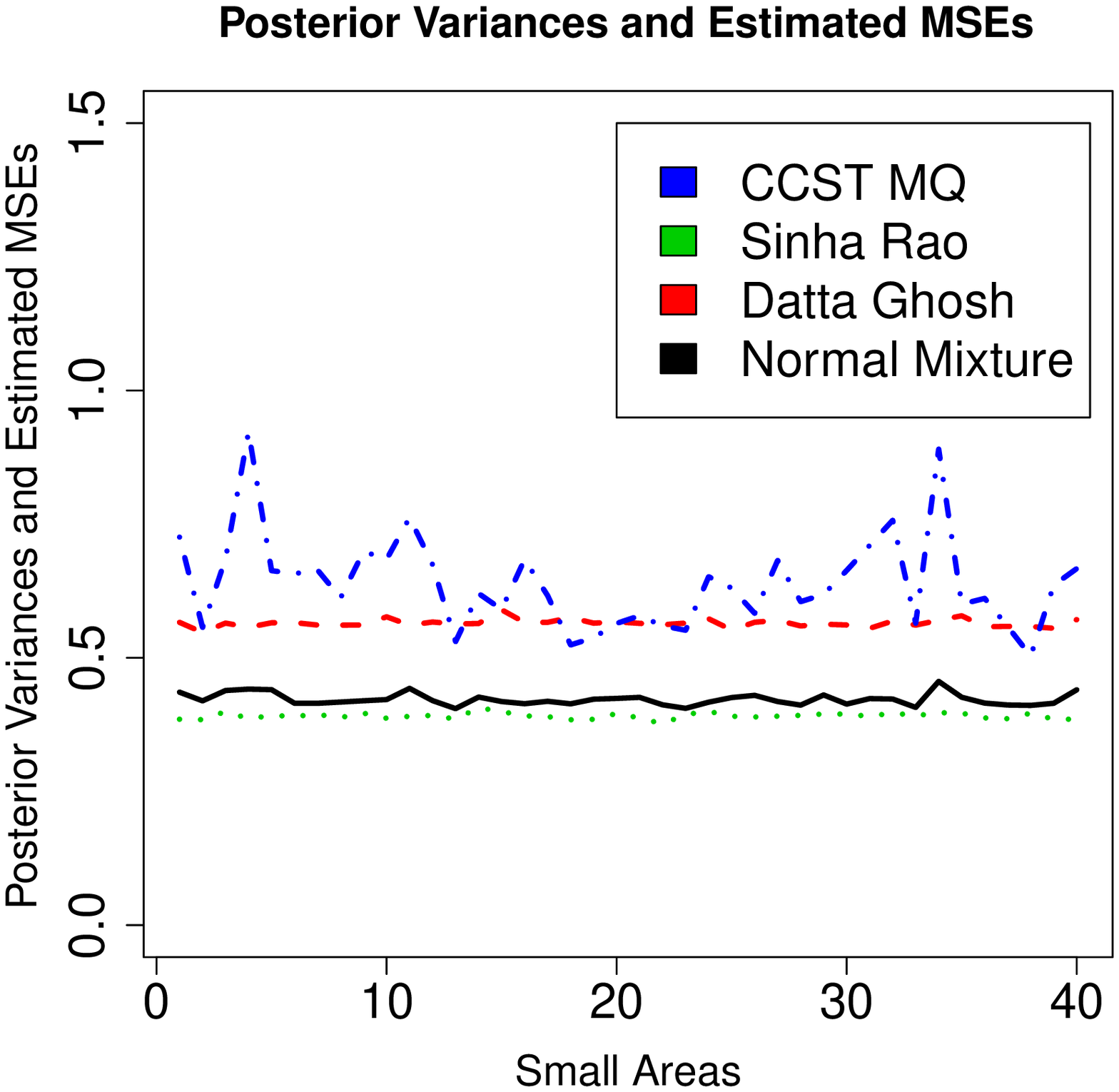} & \includegraphics[scale=.375] {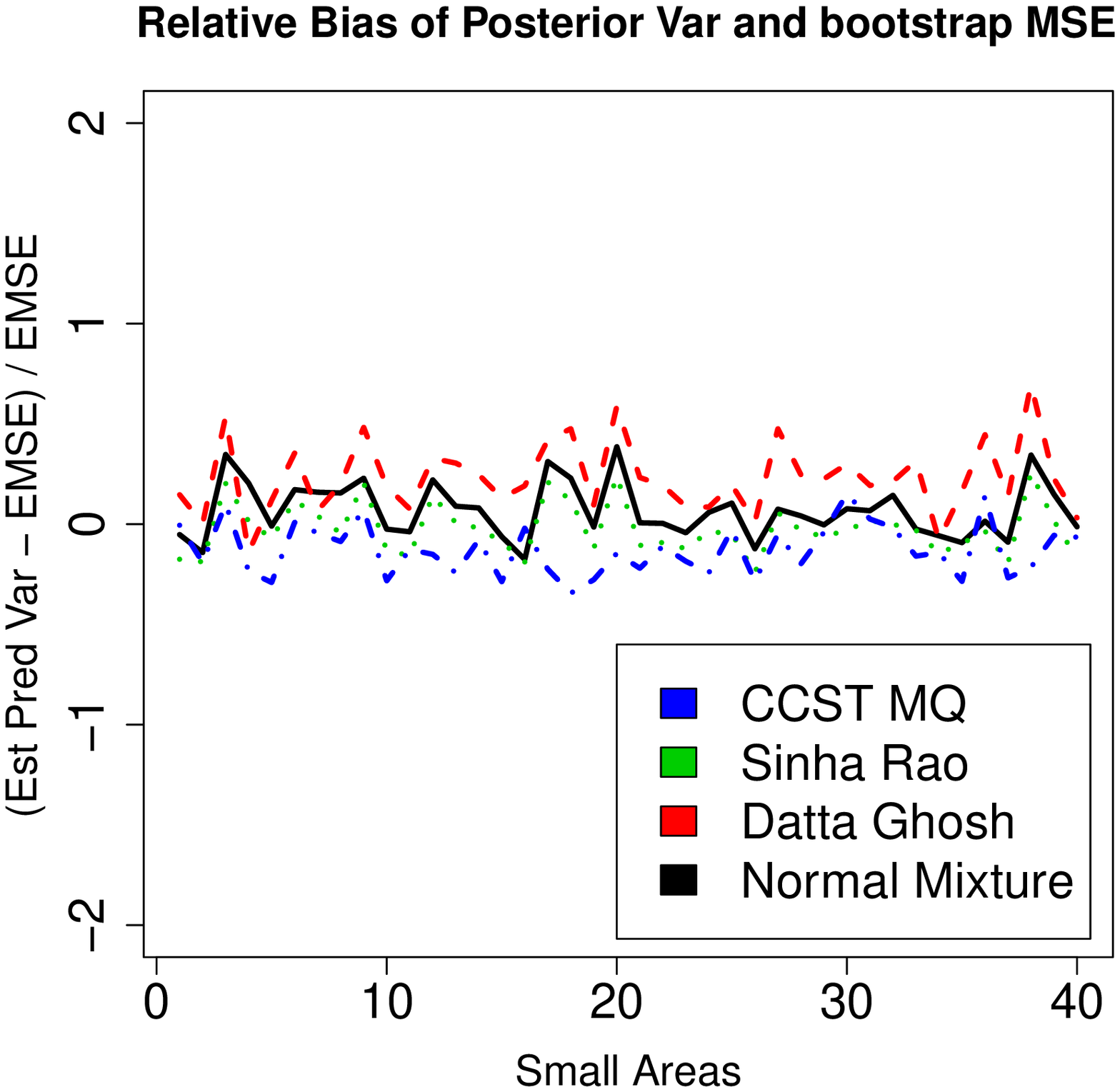}   \\
			\raisebox{3cm}{No outlier}               & \includegraphics[scale=.375]{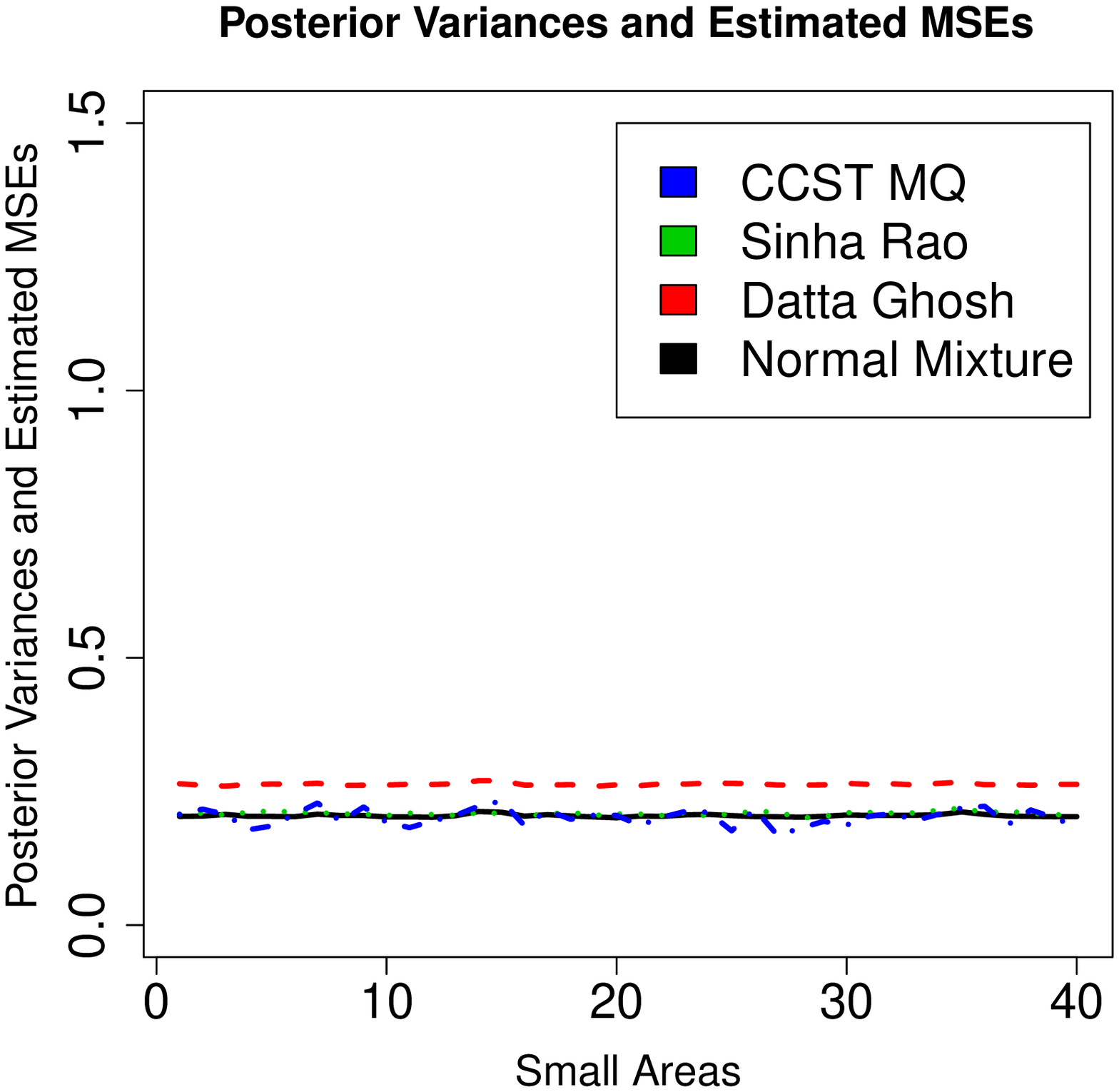}  & \includegraphics[scale=.375]{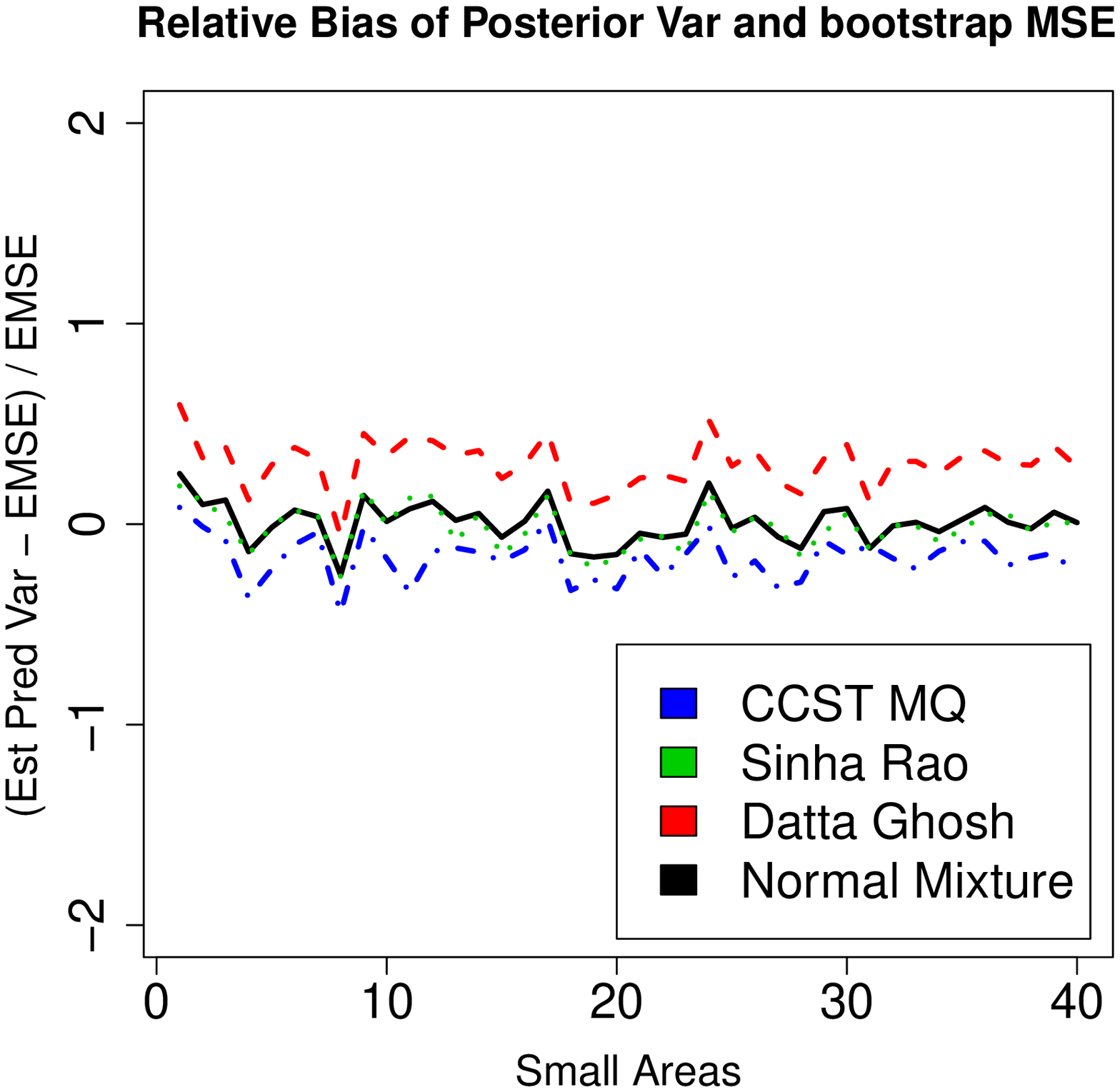}
		\end{tabular}
		\caption{Plot of posterior variances and MSE estimates and their empirical relative biases  }\label{tout:EMSE4}
	\end{center}
\end{figure}

\begin{figure}[h]
	\vspace{-1.75cm}
	\begin{center}
		\begin{tabular}{ccc}
			\raisebox{3cm}{10\%\hspace{.75cm}}       & \includegraphics[scale=.375]{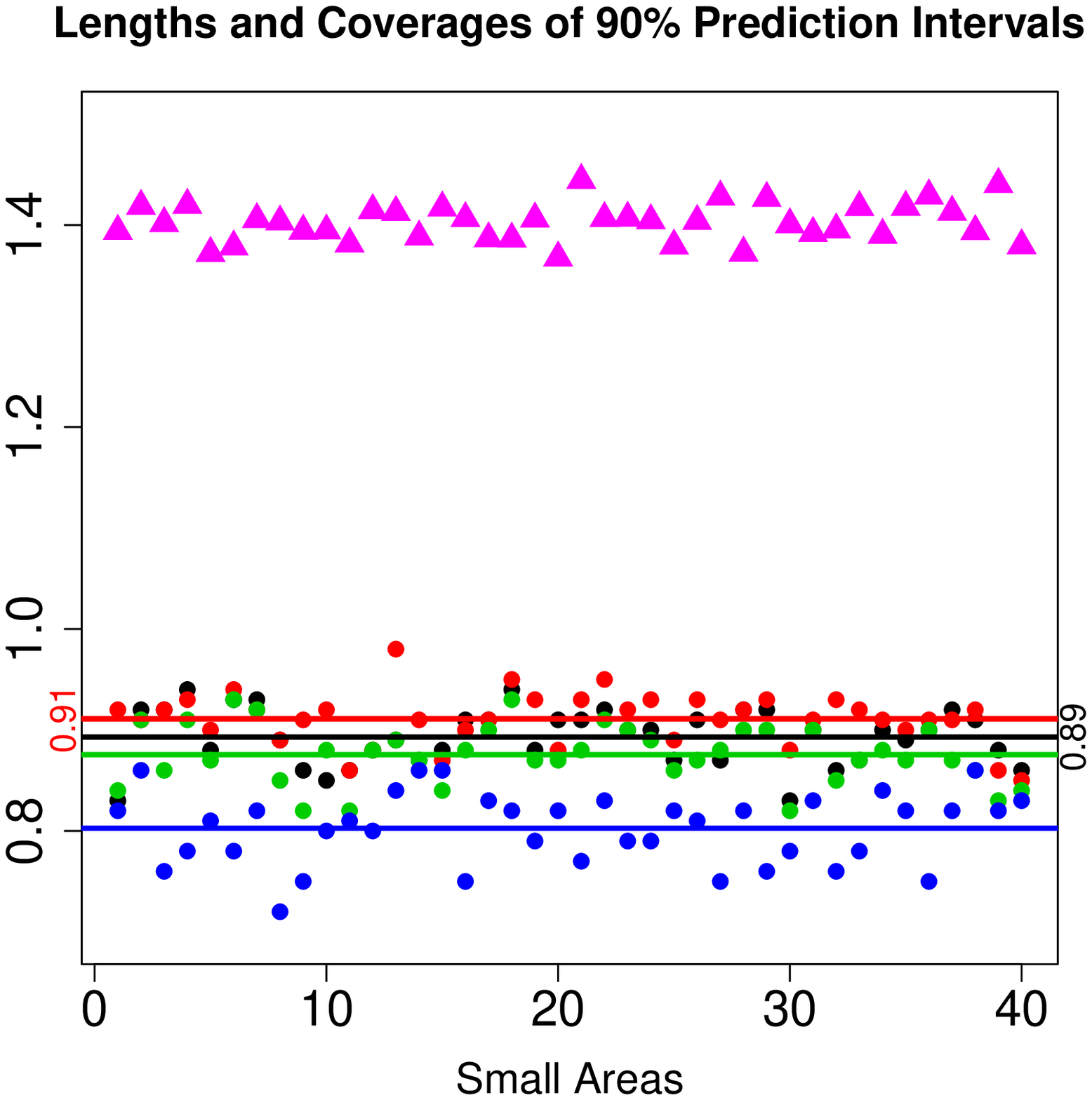} & \includegraphics[scale=.375]{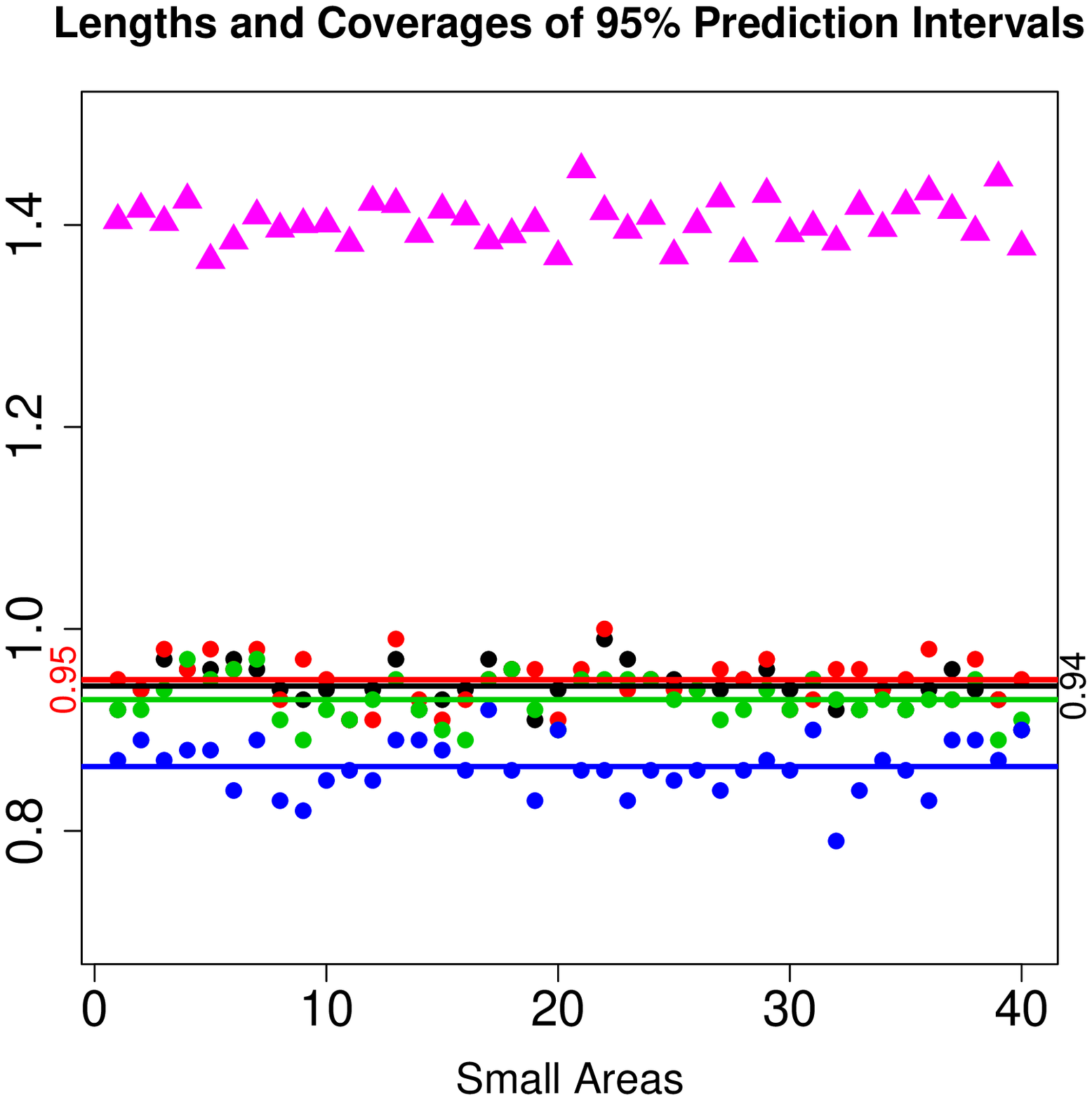} \\
			\raisebox{3cm}{$t_{(4)}$\hspace{.75cm}}  & \includegraphics[scale=.375]{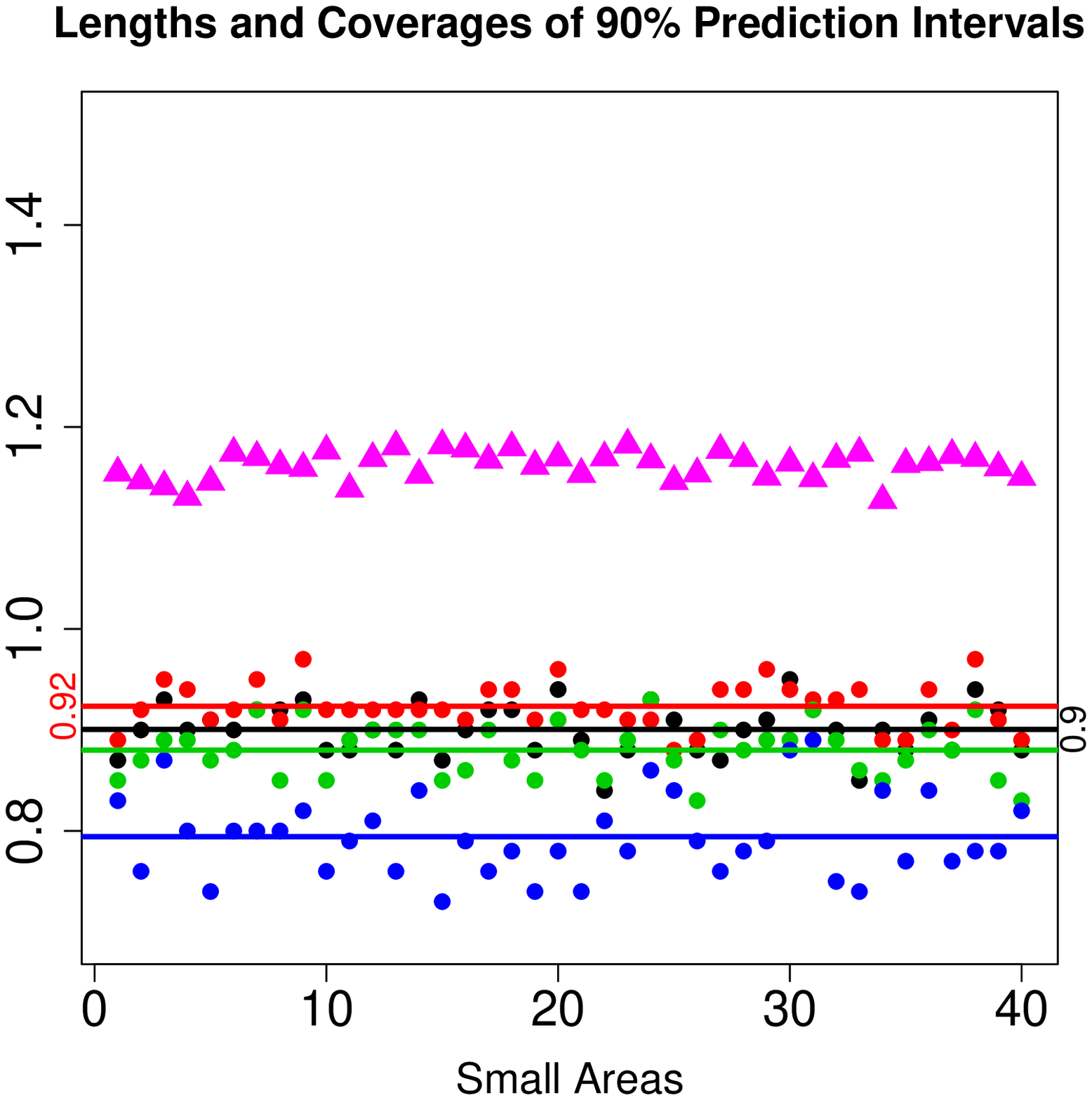}  & \includegraphics[scale=.375]{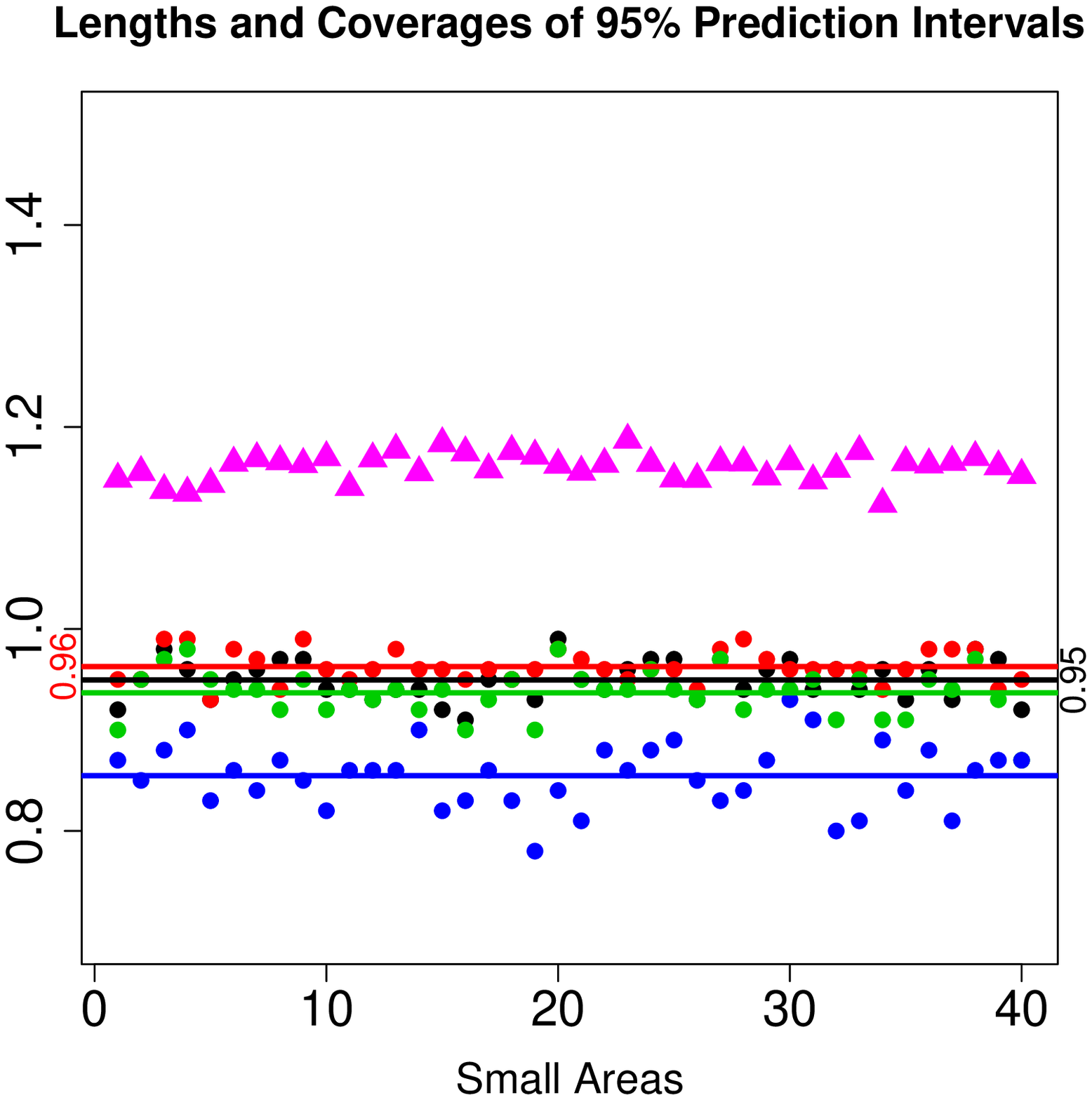}  \\
			\raisebox{3cm}{No outlier}               & \includegraphics[scale=.375]{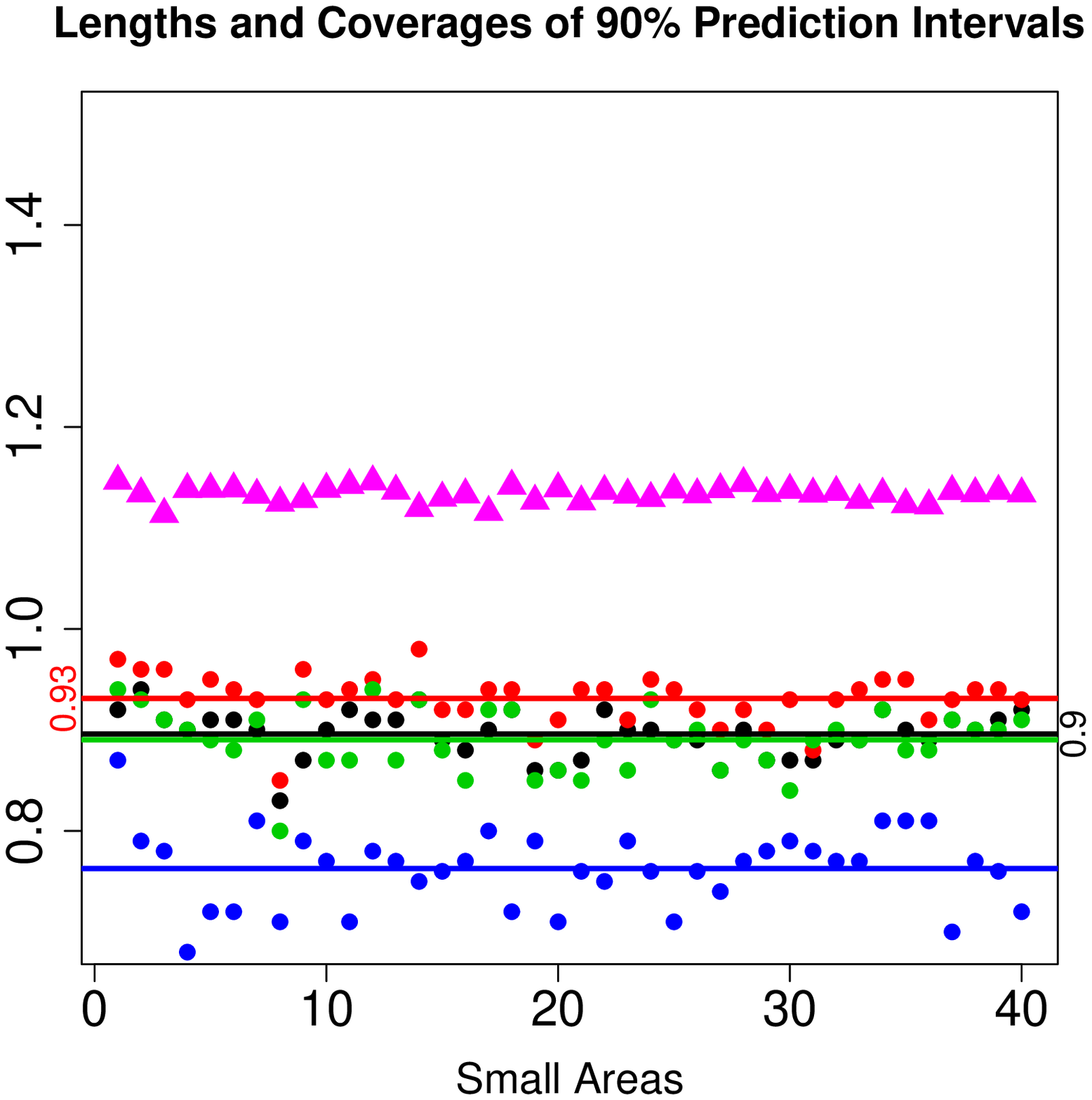} & \includegraphics[scale=.375]{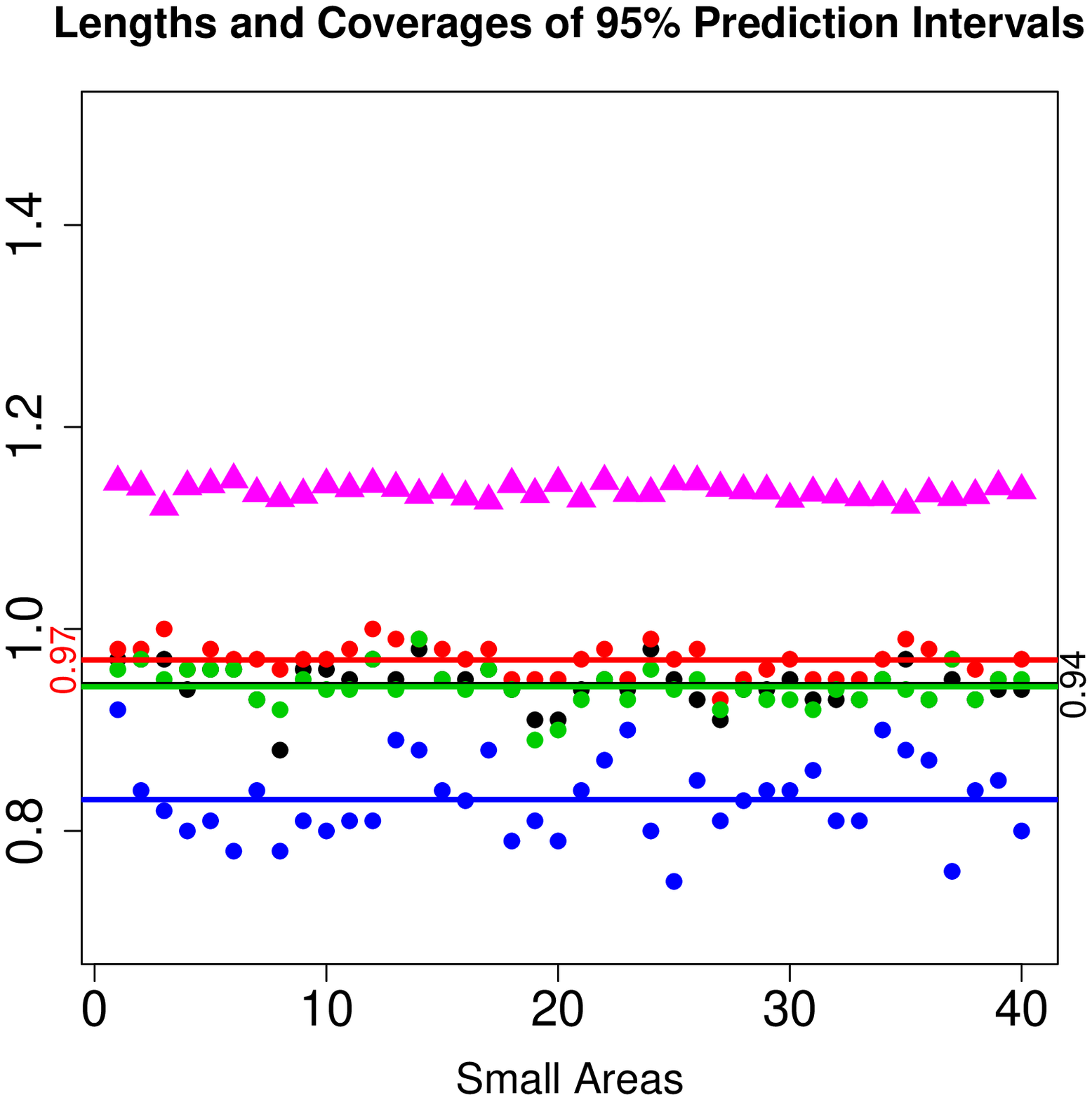} \\
		\end{tabular}
		\caption{Plot of lengths and coverages of credible and prediction intervals}\label{tout:CI2}
		\vspace{-1.075cm}
		\includegraphics[scale=.35,trim = 0 500 0 100]{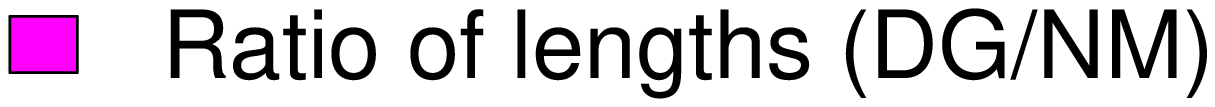} \includegraphics[scale=.35,trim = 0 500 0 010]{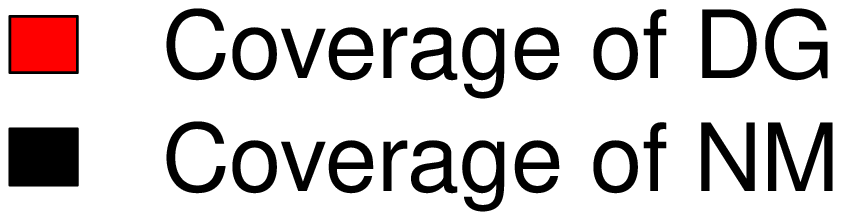} \includegraphics[scale=.35,trim = 0 500 0 010]{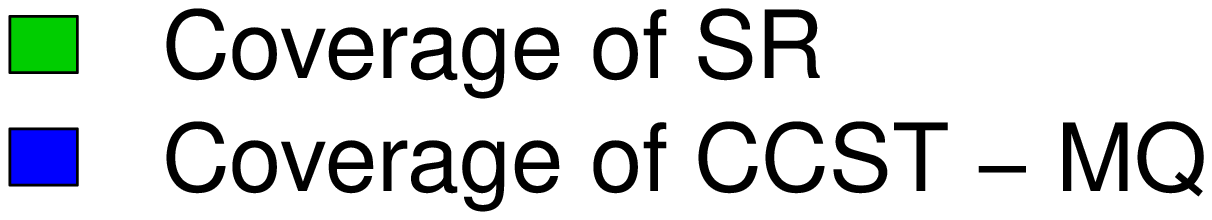}
	\end{center}
\end{figure}

\bigskip
\noindent
For each simulation setup, we have simulated $S=100$ populations. For the $s$th created population, $s=1,\cdots, S$, we computed the values of $\theta_i^{(s)}$, which will be treated as the true values. We denote the $s$th simulation sample by $d^{(s)}$, and based on this data we calculate the REBLUP predictors $\hat{\theta}_{i,SR}^{(s)}$ and their estimated MSE, $mse(\hat{\theta}_{i,SR}^{(s)})$ using the procedure proposed by Sinha and Rao (2009).
To assess the accuracy of the point predictors we computed the empirical bias $eB_{i,SR}=\frac 1S \sum_{s=1}^S (\hat{\theta}_{i,SR}^{(s)} - \theta_i^{(s)})$ and empirical MSE $eM_{i,SR}=\frac 1S \sum_{s=1}^S (\hat{\theta}_{i,SR}^{(s)} - \theta_i^{(s)})^2$. Treating $eM_{i,SR}$ as the ``true'' measure of variability of $\hat{\theta}_{i,SR}$, we also evaluate the accuracy of the MSE estimator $mse(\hat{\theta}_{i,SR})$, suggested by Sinha and Rao (2009). Accuracy of the MSE  estimator is evaluated by the relative difference between the empirical MSE and the average (over simulations) estimated MSE, given by $RE_{mse-SR,i} = \{ (1/S)\sum_{s=1}^S mse(\hat{\theta}_{i,SR}^{(s)}) - eM_{i,SR} \}/ eM_{i,SR}$. Similarly, we obtained the predictors $\hat{\theta}_{i,CCST}^{(s)}$, estimated MSEs $mse(\hat{\theta}_{i,CCST}^{(s)})$ of Chambers et al. (2014), empirical biases and empirical MSEs of point estimators and relative biases of the estimated MSEs. Using the point estimates and MSE estimates we created approximate 90\% prediction intervals $I_{i,SR,90}^{(s)} =  [\hat{\theta}_{i,SR}^{(s)} -1.645\sqrt{mse(\hat{\theta}_{i,SR}^{(s)})},\hat{\theta}_{i,SR}^{(s)} +1.645\sqrt{mse(\hat{\theta}_{i,SR}^{(s)})}] $ and 95\% prediction intervals $I_{i,SR,95}^{(s)} =  [\hat{\theta}_{i,SR}^{(s)} -1.96\sqrt{mse(\hat{\theta}_{i,SR}^{(s)})},\hat{\theta}_{i,SR}^{(s)} +1.96\sqrt{mse(\hat{\theta}_{i,SR}^{(s)})}] $. We also obtained similar intervals for the MQ method of Chambers et al. (2014).
We evaluated empirical biases, empirical MSEs, relative biases of estimated MSEs, and empirical coverage probabilities of prediction intervals for all four methods. These quantities for all 40 small areas are plotted in Figures  \ref{tout:BiasVar3}, \ref{tout:EMSE4} and \ref{tout:CI2}.

\bigskip
\noindent
We plotted the empirical biases on the left panel and the empirical MSEs on the right panel of Figure \ref{tout:BiasVar3}. These estimators do not show any systematic bias. In terms of {\it eM}, the REBLUP and the proposed NM HB predictor appear to be most accurate and perform similarly (in fact, based on all evaluation criteria considered here, the proposed NM HB and the REBLUP methods have equivalent performance). In terms of {\it eM}, the MQ predictor has maximum variability and the standard DG HB predictor is in third place. In the case of no outliers, while the other three predictors have the same {\it eM}, the MQ predictor is slightly more variable. Moreover, we examined how closely the posterior variances of the Bayesian predictors and the MSE estimators of the frequentist robust predictors track their respective {\it eM} of prediction (see Figure \ref{tout:EMSE4}). The posterior variance of the proposed NM HB predictor and the estimated MSE of REBLUP appear to track the {\it eM} the best without any evidence of bias. The posterior variance of the standard HB predictor appears to overestimate the {\it eM} and the estimated MSE of the MQ predictor appears to underestimate. An undesirable consequence of this negative bias of the MSE estimator of the MQ method is that the related prediction intervals often fail to cover the true small area means (see the plots in Figure \ref{tout:CI2}). 

\bigskip
\noindent
Our sampling-based Bayesian approach allowed us to create credible intervals for the small area means at the nominal levels of 0.90 and 0.95 based on sample quantiles of the Gibbs samples of the $\theta_i$'s. For the Sinha-Rao and the Chambers et al. methods we used their respective estimated root MSE of the REBLUPs or MQ-predictors to create symmetric approximate 90\% and 95\% prediction intervals of the small area means.

\bigskip
\noindent
To assess the coverage rate of these prediction intervals we computed empirical coverage probabilities $eC_{i,SR,90}=\frac 1S \sum_{s=1}^S I[\theta_i^{(s)} \in I_{i,SR,90}^{(s)}]$ and  $eC_{i,SR,95}=\frac 1S \sum_{s=1}^S I[\theta_i^{(s)} \in I_{i,SR,95}^{(s)}]$, where $I[x\in A]$ is the usual indicator function that is one for $x \in A$ and 0 otherwise.

\bigskip
\noindent
Based on the same setup and same set of simulated data we also evaluated the two HB procedures. In the Bayesian approach, the point predictor, the posterior variance and the credible intervals for $\theta_i^{(s)}$ in the $s$th simulation were computed based on the MCMC samples of $\theta_i^{(s)}$ from its posterior distribution, generated by Gibbs sampling. The posterior mean and posterior variance are computed by the sample mean and the sample variance of the MCMC samples. An equi-tailed $100(1-2\alpha)\%$ credible interval for $\theta_i^{(s)}$ is created, where the lower limit is the $100\alpha$th sample percentile and the upper limit is the $100(1-\alpha)$th sample percentile of the MCMC samples of $\theta_i^{(s)}$ from the $s$th simulation.

\bigskip
\noindent
Suppose in the $s$th simulation $\hat{\theta}_{i,DG}^{(s)}$ denotes the Datta-Ghosh HB predictor of $\theta_i$ and $V_{i,DG}^{(s)}$ denotes the posterior variance. The empirical bias of the Datta-Ghosh predictor of $\theta_i$ is defined by $eB_{i,DG}=\frac 1S \sum_{s=1}^S (\hat{\theta}_{i,DG}^{(s)} - \theta_i^{(s)})$ and empirical MSE by $eM_{i,DG}=\frac 1S \sum_{s=1}^S (\hat{\theta}_{i,DG}^{(s)} - \theta_i^{(s)})^2$.  To investigate the extent $V_{i,DG}^{(s)}$ may be interpreted as an estimated mse of the predictor $\hat{\theta}_{i,DG}$, we compute the relative  difference between the empirical MSE and the average (over simulations) posterior variance, given by $RE_{V-DG,i} = \{ (1/S)\sum_{s=1}^S V_{i,DG}^{(s)} - eM_{i,DG} \}/ eM_{i,DG}$. These quantities for all 40 small areas are plotted in Figure \ref{tout:EMSE4}.

\bigskip
\noindent
Based on the MCMC samples of $\theta_i$'s for the $s$th simulated data set, let $I_{i,DG,90}^{(s)}$ be the 90\% credible interval for $\theta_i$.  To evaluate the frequentist coverage probability of the credible interval for $\theta_i$ we computed empirical coverage probabilities $eC_{i,DG,90}=\frac 1S \sum_{s=1}^S I[\theta_i^{(s)} \in I_{i,DG,90}^{(s)}]$.
Corresponding to a credible interval $I_{i,DG,90}^{(s)}$, we use $L_{i,DG,90}^{(s)}$ to denote its length, and computed empirical average length of a 90\% credible interval for $\theta_i$ based on Datta-Ghosh approach by ${\bar L}_{i,DG,90} = \frac 1S \sum_{s=1}^S L_{i,DG,90}^{(s)}$. Similarly, we computed $eC_{i,DG,95}$ and ${\bar L}_{i,DG,95}$ for the 95\% credible intervals for $\theta_i$.

\bigskip
\noindent
Finally, as we did for the Datta-Ghosh HB predictor, we computed similar quantities for our new robust HB predictor. Specifically, suppose $\hat{\theta}_{i,NM}^{(s)}$ is the newly proposed NM HB predictor of $\theta_i^{(s)}$ and $V_{i,NM}^{(s)}$ is the posterior variance. For the new predictor we define the empirical bias by $eB_{i,NM}=\frac 1S \sum_{s=1}^S (\hat{\theta}_{i,NM}^{(s)} - \theta_i^{(s)})$ and empirical MSE by $eM_{i,NM}=\frac 1S \sum_{s=1}^S (\hat{\theta}_{i,NM}^{(s)} - \theta_i^{(s)})^2$.  Again, to investigate the extent $V_{i,NM}^{(s)}$ may be viewed as an estimated MSE of the predictor $\hat{\theta}_{i,NM}$, we computed the relative difference between the emprical MSE and the average (over simulations) posterior variance, given by $RE_{V-NM,i} = \{ (1/S)\sum_{s=1}^S V_{i,NM}^{(s)} - eM_{i,NM} \}/ eM_{i,NM}$. These quantities for all 40 small areas are plotted in Figure  \ref{tout:EMSE4}. Based on the MCMC samples of $\theta_i$'s for the $s$th simulated data set, let $I_{i,NM,90}^{(s)}$ be the 90\% credible interval for $\theta_i$.  To evaluate the frequentist coverage probability of the credible interval for $\theta_i$ we computed empirical coverage probabilities $eC_{i,NM,90}=\frac 1S \sum_{s=1}^S I[\theta_i^{(s)} \in I_{i,NM,90}^{(s)}]$.
Corresponding to a credible interval $I_{i,NM,90}^{(s)}$, we use $L_{i,NM,90}^{(s)}$ to denote its length, and computed empirical average length of a 90\% credible interval for $\theta_i$ based on new approach by ${\bar L}_{i,NM,90} = \frac 1S \sum_{s=1}^S L_{i,NM,90}^{(s)}$. Similarly, we computed $eC_{i,NM,95}$ and ${\bar L}_{i,NM,95}$ for the 95\% credible intervals for $\theta_i$.

\bigskip
\noindent
We plotted the empirical coverage probabilities for the four methods that we considered in this article. The plot reveals significant undercoverage of the approximate prediction intervals created by using the estimated prediction MSE proposed by Chambers et al. (2014). This undercoverage is not surprising since their estimated MSE mostly underestimates the true MSE (measured by the {\it eM}) (see  Figure \ref{tout:EMSE4}). Coverage probabilities of the Sinha-Rao prediction intervals and the two Bayesian credible intervals are remarkably accurate. This lends dual interpretation of our proposed credible intervals, Bayesian by construction, and frequentist by simulation validation. This property is highly desirable to practitioners, who often do not care about a paradigm or a philosophy. In the same plot, we also plotted the ratio of the average lengths of the DG credible intervals to the newly proposed robust HB credible intervals. These plots show the superiority of the proposed method, yielding intervals which meet coverage accurately  with average lengths about 25-30\% shorter compared to the DG method for normal mixture model with 10\% contamination. Again these two intervals meet the coverage accurately when the unit-level errors are generated from normal (no outliers) or a moderately heavy-tail distribution ($t_4$). In these cases, the reduction in length of the intervals is less, which is about 10\%. This shorter prediction intervals from the new method even for normal distribution for the unit-level error is interesting; it shows that the proposed method does not lose any efficiency in comparison with the Datta-Ghosh method even when the normality of the unit-level errors holds.

\bigskip
\noindent
The comparison of NM HB prediction intervals and the Sinha-Rao prediction intervals yields a mixed picture. In the mixture setup, the NM HB prediction intervals attained coverage probability more accurately than the Sinha-Rao intervals, which undercover by 1\%, and on an average the Bayesian prediction intervals are about 2\% shorter than the frequentist intervals. When the data are simulated from a $t_4$ distribution, the coverage probabilities of the Sinha-Rao prediction intervals are about 1\% below the target, but these intervals are about 3\% shorter than the NM HB prediction intervals, which attained the nominal coverage. Finally, when the population does not include any outlier, these two methods perform the same, both attained the nominal coverage and yield the same average length.

\section{Conclusion}
The NER model by Battese et al. (1988) plays an important role in small area estimation for unit-level data. While Battese et al. (1988), Prasad and Rao (1990) and Datta and Lahiri (2000) investigated EBLUPs of small area means, Datta and Ghosh (1991) proposed an HB approach for this model. Sinha and Rao (2009) investigated robustness of the MSE estimates of EBLUPs in Prasad and Rao (1990) for outliers in the response. They showed in presence of outliers robustness of their REBLUPs and lack of robustness of the EBLUPs.

\bigskip
\noindent
In this article we showed that non-robustness also persists for the HB predictors by Datta and Ghosh (1991). To deal with this undesirable issue we proposed an alternative to the HB predictors by using a mixture of normal distributions for the unit-level error part of the NER model.  An illustrative application and simulation study show the superiority of our proposed method over the existing HB, EBLUP and M-quantile solutions. Indeed simulation results show the superiority of our method over the Datta and Ghosh (1991) HB predictors and the M-quantile small area estimators of Chambers et al. (2014). Performance of our proposed NM HB method is found to be as good as the frequentist solution of Sinha and Rao (2009). Our proposed Bayesian intervals also achieve the corresponding frequentist coverage. Thus, unlike the frequentist solutions, our proposed HB solution enjoys dual interpretation, Bayesian by construction, and frequentist via simulation, a feature attractive to practitioners. Moreover, suggested credible intervals are shorter in length in comparison with the other nominal prediction intervals. In fact,  the application and simulations show the proposed NM HB method is the best among the four methods in presence of outliers. Our proposed method is as good as the HB method of Datta and Ghosh (1991), even in absence of outliers. Thus there will be no loss in using the proposed HB method for all data sets.  It is not clear to us that why M-quantile performs poorly. However, we note  that in our simulations, all the errors are centered at zero. Alternatively, one can explore the performance of these methods when the outlier parts of the respective error components are generated from a distribution which is not centered at zero. This remains a topic of future research.

\section{Acknowledgment}

Authors are thankful to Drs. Bill Bell and Jerry Maples for their insightful comments.

\section*{Supporting Information}

Property of the posterior distribution corresponding to the proposed 
model has been discussed in the supplementary material.

\clearpage

\section*{References}

\def\beginref{\begingroup
                \clubpenalty=10000
                \widowpenalty=10000
                \normalbaselines\parindent 0pt
                \parskip.0\baselineskip
                \everypar{\hangindent1em}}
\def\endref{\par\endgroup}

\beginref

Battese, G. E., Harter, R. M. and Fuller, W. A. (1988), An error component model for prediction of county crop areas using survey and satellite data, {\em Journal of the American Statistical Association}, {\bf 83}, 28--36.

Bell, W. R. and Huang, E. T. (2006). Using the $t$-distribution to deal with outliers in small area estimation. In {\it Proceedings of Statistics Canada Symposium on Methodological Issues in Measuring Population Health}. Statistics Canada, Ottawa, Canada.

Breckling, J. and Chambers, R. (1988), M-quantiles, {\it Biometrika}, {\bf 75}, $761-771$


Chakraborty, A., Datta, G. S. and Mandal, A. (2016), A two-component normal mixture alternative to the Fay-Herriot model, {\em Statistics in Transition new series and Survey Methodology Joint Issue: Small Area Estimation 2014}, {\bf 17},  67--90.

Chambers, R. L. (1986), Outlier robust finite population estimation, {\em Journal of the American Statistical Association}, {\bf 81}, 1063--1069.

Chambers, R., Chandra, H., Salvati, N. and Tzavidis, N. (2014), Outlier robust small area estimation. {\em Journal of the Royal Statistical Society Series B}, {\bf 76}, 47–-69.

Chambers, R.L. and Tzavidis, N. (2006), M-quantile models for small area estimation, {\it Biometrika}, {\bf 93}, $255-268.$.

Datta, G. and Ghosh, M. (1991), Bayesian prediction in linear models: Applications to small area estimation, {\em Annals of Statistics}, {\bf 19}, 1748--1770.

Datta, G. S. and Lahiri, P. (1995), Robust hierarchical Bayesian estimation of small area characteristics in presence of covariates and outliers,  {\em Journal of Multivariate Analysis}, {\bf 54}, 310--328.

Datta, G. S. and Lahiri, P. (2000), A unified measure of uncertainty of estimated best linear unbiased predictors in small area estimation problems, {\em Statistica Sinica}, {\bf 10}, 613--627.

Datta, G. S.,  Rao,  J. N. K. and Smith, D. D. (2005), On measuring the variability of small area estimators under a basic area level model, {\em  Biometrika}, {\bf 92}, 183--196.

Fay, R. E. and Herriot, R. A. (1979), Estimates of income for small places:
an application of James-Stein procedures to census data, {\em Journal of the American Statistical Association}, {\bf 74}, 269--277.

Fellner, W. H.  (1986), Robust estimation of variance components, {\em  Technometrics}, {\bf 28}, 51--60.

Fabrizi, E., Salvati, N. and Pratesi (2012), M. Constrained small area estimators based on M-quantile methods. {\em Journal of Official Statistics}, {\bf 28}, 89–-106.

Gershunskaya, J. (2010). Robust Small Area Estimation Using a Mixture Model. {\it Proceedings of the Section on Survey Research Methods Section},  Washington, DC: American Statistical Association. 

Hobert, J. and Casella, G. (1996), Effect of improper priors on Gibbs sampling in hierarchical linear mixed models, {\em Journal of the American Statistical Association}, {\bf 91}, 1461--1473.

Lahiri, P. and Rao, J.N.K. (1995), Robust estimation of mean square error of small area estimators, {\em Journal of the American Statistical Association}, {\bf 90}, 758--766.

Pfeffermann, D. and Sverchkov, M. (2007), Small area estimation under informative probability sampling of areas and within the selected areas, {\em Journal of the American Statistical Association}, {\bf 102}, 1427--1439.

Prasad, N. G. N. and Rao, J. N. K. (1990), On the estimation of mean square error of small area predictors,  {\em Journal of the American Statistical Association}, {\bf 85}, 163--171.

Sinha, S. K. and Rao, J. N. K. (2009), Robust small area estimation, {\em The Canadian Journal of Statistics}, {\bf 37}, 381--399.

{Tzavidis, N. and Chambers, R. (2005), Bias adjusted estimation for small areas with M-quantile models,
{\em Statistics in Transition} {\bf 7}, 707--713.}

Verret, F., Rao, J. N. K. and Hidiroglou, M. (2015), Model-based small area estimation under informative sampling, {\em Survey Methodology}, {\bf 41}, 333--347.
\endref

\clearpage

\clearpage

\clearpage

\end{document}